\documentclass[11pt]{article}

\usepackage{amsmath}
\usepackage{bm}
\usepackage{amssymb}
\usepackage{amstext}
\usepackage{latexsym}
\usepackage{color}
\usepackage{graphicx}
\usepackage{epsfig}
\usepackage{subfigure}
\usepackage{rotating}
\usepackage{curves}
\usepackage{epic}
\usepackage{amsthm}
\allowdisplaybreaks[1]

\usepackage{amsfonts}
\usepackage[T1]{fontenc}
\usepackage[latin1]{inputenc}
\usepackage{authblk}
\usepackage{tikz}
\usetikzlibrary{shapes,arrows}
\usepackage{graphics}
\usepackage{verbatim}
\usepackage{color}
\usepackage{hyperref}
\usepackage{mathtools}

\newcommand{\be}{\begin{equation}}
\newcommand{\en}{\end{equation}}

\newtheorem{thm}{Theorem}
\newtheorem{cor}[thm]{Corollary}

\newtheorem{defi}{Definition}[section]
\newtheorem{lem}[defi]{Lemma}

\newtheorem{Theo}{Theorem}[section]

\newtheorem{remark}[Theo]{Remark}

\newcommand{\bedefin}{\begin{defi}}
\newcommand{\findefi}{\end{defi} \medskip}

\newcommand{\betheo}{\begin{theorem}$\!\!${\bf \,\,\,}}
\newcommand{\entheo}{\end{theorem}}
\newcommand{\enth}{\end{theorem}}

\newcommand{\becor}{\begin{cor}$\!\!${\bf .}}
\newcommand{\encor}{\end{cor}}

\newcommand{\belem}{\begin{lem}$\!\!${\bf }}
\newcommand{\enlem}{\end{lem}}

\newcommand{\prf}{\noindent{\bf{ Proof}\,\,}}

\newcommand{\bea}{\begin{eqnarray}}
\newcommand{\ena}{\end{eqnarray}}

\newcommand{\beano}{\begin{eqnarray*}}
\newcommand{\enano}{\end{eqnarray*}}

\newcommand{\bee}{\begin{enumerate}}
\newcommand{\ene}{\end{enumerate}}

\newcommand{\bei}{\begin{itemize}}
\newcommand{\eni}{\end{itemize}}

\newcommand{\betab}{\begin{tabular}}
\newcommand{\entab}{\end{tabular}}

\newcommand{\bd}{\begin{displaymath}}

\newcommand{\nn}{\nonumber}

\newcommand{\bp}{\mathbf p}
\newcommand{\bq}{\mathbf q}
\newcommand{\br}{\mathbf r}

\newcommand{\bn}{\mathbf n}
\newcommand{\bx}{\mathbf x}
\newcommand{\bs}{\mathbf s}

\newcommand{\bA}{\mathbf A}

\newcommand{\g}{G_{\hbox{\tiny{NC}}}}
\newcommand{\gd}{\hat{G}_{\hbox{\tiny{NC}}}}
\newcommand{\G}{\mathfrak{g}_{\hbox{\tiny{NC}}}}
\newcommand{\gwh}{G_{\hbox{\tiny{WH}}}}
\newcommand{\GWH}{\mathfrak{g}_{\hbox{\tiny{WH}}}}

\catcode `\@=11 \@addtoreset{equation}{section}

\catcode `\@=12

\begin{document}

\title{On the Plethora of Representations Arising in Noncommutative Quantum Mechanics and An Explicit Construction of Noncommutative 4-tori}
\author[1,2]{S. Hasibul Hassan Chowdhury\thanks{shhchowdhury@gmail.com}}
\affil[1]{Chern Institute of Mathematics, Nankai University, Tianjin 300071, P. R. China}
\affil[2]{Laboratory of Computational Sciences and Mathematical Physics, Institute for
Mathematical Research, Universiti Putra Malaysia, 43400 UPM Serdang, Selangor,
Malaysia}

\date{\today}

\maketitle

\begin{abstract}
We construct a 2-parameter family of unitarily equivalent irreducible representations of the triply extended group $\g$ of translations of $\mathbb{R}^{4}$ associated with a family of its 4-dimensional coadjoint orbits and show how a continuous 2-parameter family of gauge potentials emerges from these unitarly equivalent representations. We show that the Landau and the symmetric gauges of noncommutative quantum mechanics, widely used in the literature, in fact, belong to this 2-parameter family of gauges. We also provide an explicit construction of noncommutative 4-tori and compute the associated star products using the unitary dual of the group $\g$ that was studied at length in an earlier paper (\cite{ncqmjpa}). Finally, we construct projective modules over such noncommutative 4-tori and compute constant curvature connections on them using Rieffel's method.
\end{abstract}

\section{Introduction}\label{sec:intro}
It has long been argued that geometry of space-time should be modified to accommodate spatial noncommutativity at lengths as small as Planck length. Among others, Snyder and Yang were the pioneers to investigate such noncommutative structure of space-time (see \cite{Sny,Yang}). Spatial localization at an arbitrarily large accuracy can lead to possible creation of black holes contributing to the loss of operational meaning of space-time as has been argued by Doplicher et al. in \cite{Doplicheretal}. Motivated by these arguments, one can then consider a noncommutative phase-space where in addition to the quantum mechanical position-momentum noncommutativity, one incorporates noncommutativity between the operators representing spatial coordinates. Quantum mechanics in noncommutative phase-space is generally referred to as {\em noncommutative quantum mechanics}. It is abbreviated as NCQM in the sequel. Phase-space formulation of quantum mechanics has been introduced in various articles (see, for example, \cite{Torre-Vegaetal} and many articles cited therein). Quantum mechanics on noncommutative phase-space, on the other hand, has started drawing attention of the physicists very recently (see, for example \cite{Muth,bertolamietal}).

Numerous articles were written, of late, delineating formulations and applications of NCQM in Physics ranging from solid state Physics to string theory (see \cite{berardetal,scholtzt,balachandranetal}). The most widely advocated physical applications of NCQM is provided by quantum mechanical systems coupled to a constant background magnetic field. Refer to \cite{jackiw} for a detailed account on the relevant physical applications along this line with some historical background.

NCQM has also been considered as being non-relativistic approximation of noncommutative quantum field theory (NCQFT) (see \cite{Hoetal}) where the underlying fields are considered as functions of a noncommutative space-time with spatial coordinates failing to commute with each other. A detailed account on the modern aspects of noncommutative quantum field theory can be found in \cite{Douglasetal,Szabo}.

In a group theoretic formulation of NCQM for a system of $2$ degrees of freedom, the authors in \cite{ncqmjmp,ncqmjpa} start with a connected, simply connected nilpotent Lie group and obtain various unitary irreducible representations of $\g$ and its Lie algebra $\G$ following the method of orbits (see \cite{Kirillovbook}). This nilpotent Lie group was later identified with the kinematical symmetry group for this model of NCQM with $2$ degrees of freedom in \cite{wigfuncpaper} by computing its various Wigner functions supported on the respective coadjoint orbits. The Wigner functions, thus constructed, are then verified to agree with the quantum mechanical Wigner functions, originally computed by Wigner in his seminal paper \cite{wig}.

We digress a bit on differential geometric and $C^{*}$-algebraic setting by roughly sketching some basic constructs of {\em noncommutative geometry}. Here, one starts with the commutative $C^{*}$-algebra of smooth functions on a compact Hausdorff space $\mathcal{X}$ and replace it with a noncommutative $C^{*}$-algebra of operators defined on an infinite dimensional Hilbert space with the commutative point-wise product of $C^{\infty}(\mathcal{X})$ now deformed into a noncommutative product. One is also required to ensure in this case that under appropriate limit the commutative $C^{*}$-algebra $C^{\infty}(\mathcal{X})$ is recovered from the underlying noncommutative $C^{*}$-algebra of operators on the respective Hilbert space. Now in ordinary (commutative) differential geometry, one constructs finite rank vector bundles over $\mathcal{X}$ and compute connections on them. Analogous construction can be achieved in the setting of noncommutative $C^{*}$-algebras motivated by Serre-Swan theorem (see, for example, \cite{Garcia-Bondiaetal}) which states that the category of vector bundles over $\mathcal{X}$ is equivalent to that of finitely generated projective modules over $C^{\infty}(\mathcal{X})$. The objects of the later category are nothing but the space of smooth sections of vector bundle over $\mathcal{X}$. The concept of vector bundles in the case of noncommutative $C^{*}$-algebras can then be generalized as the finitely generated projective modules over them. One can then proceed to suitably define connections on such projective modules over noncommutative $C^{*}$-algebras and compute the relevant curvatures.

A multitude of articles (see, for example, \cite{delducetalt,dulat}) were written of late studying Landau problem in NCQM using magnetic vector potentials in both {\em Landau} and {\em symmetric gauges}. The first aim of this paper is to compute explicitly a continuous family of gauges arising in the framework of NCQM and show in particular that the most frequently used gauges in this context, i.e. the Landau and the symmetric gauges can be obtained from this family of gauges by fixing the values of the underlying continuous parameters. This continuous family of gauges, in turn, can be shown to directly follow from a 2-parameter continuous family of equivalent unitary irreducible representations of the kinematical symmetry group $\g$ of NCQM. The second aim of the paper is to construct a noncommutative 4-tori (refer to \cite{rieffel} or section \ref{sec:NC-tori-NCQM} for definition) explicitly using the various continuous families of unitary irreducible representations of $\g$ and study noncommutative geometry on such noncommutative space. In particular, we construct the finitely generated projective modules over such noncommutative 4-tori and define connections of constant curvature on them. In the Yang-Mills theory of noncommutative tori, connections of constant curvature arise naturally as the solution of Yang-Mills equation, i.e. they extremize the Yang-Mills functional defined on the space of connections or Yang-Mills fields (see, for example, \cite{spera,rieffel-YM} ). Constant curvature connections on projective modules over noncommutative tori are also related to $\frac{1}{2}$ BPS states in super Yang-Mills theory (see \cite{CDS}).

The organization of the paper is as follows. Section \ref{sec:gwh-vs-gnc} provides a representation theoretic comparison between the $5$-dimensional {\em Weyl-Heisenberg group} and the $7$-dimensional {\em triply extended group of translations of $\mathbb{R}^{4}$}, denoted by $\gwh$ and $\g$, respectively. In section \ref{sec:algbrc-strctr-review}, we review the algebraic structure associated with the group $\g$ and enumerate its various coadjoint orbits lying in the dual Lie algebra $\G^*$. In section \ref{sec:classfctn-rep-grp-algbr}, we recapitulate the classification of the unitary irreducible representations (UIRs) of $\g$ obtained in \cite{ncqmjpa}. In \cite{ncqmjpa}, the UIRs of $\g$ were not all computed on the configuration space. Therefore, we inverse-Fourier transform the results of \cite{ncqmjpa} and list them in section \ref{sec:classfctn-rep-grp-algbr} to facilitate the computations of the following section. Section \ref{sec:gauge-sec} is devoted to the study of the family of NCQM gauges and their relation to certain family of unitarily equivalent irreducible representations of $\g$. In section \ref{sec:NC-tori-NCQM}, we construct noncommutative 4-tori using the unitary dual $\gd$ listed in section \ref{sec:classfctn-rep-grp-algbr}. Star-product between elements of $C^{\infty}(\mathbb{T}^{4})$ is introduced in section \ref{sec:star-products}. In section \ref{sec:classfctn-proj-mod}, following the construction of projective modules over the underlying noncommutative $4$-tori, we define connections of constant curvature on them. Finally, in section \ref{sec:conclsn-perspctve}, we give our closing remarks and mention some possible future work.

\section{A comparative study between $\gwh$ and $\g$}\label{sec:gwh-vs-gnc}

The Weyl-Heisenberg group $\gwh$ in 2-dimensions, being a nilpotent Lie group defines a nonrelativistic quantum mechanical system with 2 degrees of freedom. Method of orbits due to Kirillov (see \cite{Kirillovbook}) can be employed to compute the family of unitary irreducible representations of this defining group of quantum mechanics. For each fixed value of Planck's constant, denoted by $\hbar$, one obtains an equivalence class of unitary irreducible representations of $\gwh$. 

The phase space of a nonrelativistic system of 2 degrees of freedom is 4-dimensional with 2 positions and 2 momenta coordinates. $\gwh$ is just a nontrivial central extension of the underlying Abelian group of translations in $\mathbb{R}^{4}$, a group element of which is denoted by $(q_1,q_2,p_1,p_2)$. A generic element of the 5-dimensional Lie group $\gwh$ is represented by $(\theta,q_1,q_2,p_1,p_2)$. Therefore, the underlying dual Lie algebra is also a 5-dimensional real vector space. 

There is a natural action of $\gwh$ on its dual Lie algebra called the coadjoint action. The symplectic leaves of foliation of the 5-dimensional dual Lie algebra are precisely the orbits under this coadjoint action, a.k.a. coadjoint orbits. The underlying coadjoint orbits are all 4-dimensional. These codimension 1 coadjoint orbits are parametrized by the nonzero Planck's constant $\hbar$ and each such nonzero real value of $\hbar$ corresponds to a unitary irreducible representation of the 5-dimensional Lie group $\gwh$ on $L^{2}(\mathbb{R}^{2})$. The Weyl-Heisenberg Lie algebra denoted by $\GWH$, on the other hand, admits a realization of self adjoint differential operators on the smooth vectors of $L^{2}(\mathbb{R}^{2})$, the commutation relations for which read as follows:
\begin{equation}\label{eq:CCR}
\begin{aligned}
&[\hat{Q}_1,\hat{P}_1]=[\hat{Q}_2,\hat{P}_2]=i\hbar\mathbb{I}.
\end{aligned}
\end{equation}
Here, $\hat{Q}_i$'s and $\hat{P}_i$'s are the self-adjoint representations of the Lie algebra basis elements $Q_i$'s
and $P_i$'s where $i=1,2$. Note that the noncentral basis elements $Q_i$'s  and $P_i$'s correspond to the group parameters $p_i$'s and $q_i$'s, respectively, for $i=1,2$. Also, $\mathbb{I}$ stands for the identity operator on $L^{2}(\mathbb{R}^{2})$ and the central basis element $\Theta$ of the algebra is mapped to scalar multiple of $\mathbb{I}$.

In contrast to the well-known and much studied Lie group $\gwh$, if one considers 3 inequivalent local exponents (see \cite{ncqmjmp}) of the Abelian group of translations in $\mathbb{R}^{4}$ and extend it centrally using them to obtain a 7-dimensional real Lie group $\g$, the geometry of the underlying coadjoint orbits and the pertaining theory of group representations are found to be vastly rich as studied in good detail in (\cite{ncqmjpa}). 

The aim of introducing two other inequivalent local exponents besides the one used to arrive at $\gwh$ was to incorporate position-position and momentum-momentum noncommutativity as employed in the formulation of noncommutative quantum mechanics (NCQM). It is in this sense, $\g$, is termed as the defining group of NCQM in (\cite{ncqmjpa}). A generic element of $\g$ will be denoted by $(\theta,\phi,\psi,q_1,q_2,p_1,p_2)$ where $(\theta,\phi,\psi)$ forms the 3-dimensional center of the group. The Lie algebra and the dual Lie algebra of $\g$ will be denoted by $\G$ and $\G^*$, respectively, in the sequel. They are both 7-dimensional real vector spaces. The unitary dual of $\g$, i.e. the equivalence classes of unitary irreducible representations of $\g$ is denoted by $\gd$.

If one denotes the generators of $\g$ corresponding to the group parameters $q_1$, $q_2$, $p_1$ and $p_2$ by $P_1$, $P_2$, $Q_1$ and $Q_2$, respectively, then they can be suitably realized as selfadjoint differential operators, namely, $\hat{P}_1$, $\hat{P}_2$, $\hat{Q}_1$ and $\hat{Q}_2$, respectively, on the space of smooth vectors of $L^{2}(\mathbb{R}^{2})$ obeying the following set of nonvanishing commutation relations:
\begin{equation}\label{eq:NCCR}
\begin{aligned}
&[\hat{Q}_1,\hat{P}_1]=[\hat{Q}_2,\hat{P}_2]=i\hbar\mathbb{I},\\
&[\hat{Q}_1,\hat{Q}_2]=i\vartheta\mathbb{I},\;\hbox{and}\;[\hat{P}_1,\hat{P}_2]=i\mathcal{B}\mathbb{I}.
\end{aligned}
\end{equation}
Here, the central generators associated with the group parameters $\theta$, $\phi$ and $\psi$ are all mapped to scalar multiples of the identity operator $\mathbb{I}$ on $L^{2}(\mathbb{R}^2)$. The triple $(\hbar,\vartheta,\mathcal{B})$ determines the 4-dimensional coadjoint orbit, lying in the 7-dimensional dual Lie algebra $\G^*$, to which the UIR (\ref{eq:NCCR}) of $\G$ corresponds by the method of orbit.

There is yet another interesting family of 4-dimensional coadjoint orbits lying in $\G^*$ that are parametrized by a single parameter $\hbar$. The UIRs of the Lie algebra $\G$ associated with this family of coadjoint orbits obey the canonical commutation relations (CCR) (see \ref{eq:CCR}) of quantum mechanics. Therefore, one does not need to resort to the representation theory of the 5-dimensional Lie group $\gwh$ to obtain the CCR of a nonrelativistic system in two degrees of freedom as the unitary dual $\gd$ contains the family of UIRs of $\gwh$ as its own representation. It ought to be noted in this context that $\gwh$ is not a subgroup of $\g$.

$\g$ has other families of 4-dimensional coadjoint orbits which represent unitarily inequivalent representations of the group and the commutation relations involved there are also very different from each other. In addition to the 4-dimensional ones, $\g$ admits 2-dimensional and 0-dimensional coadjoint orbits. The UIRs associated with these orbits have all been classified in (\cite{ncqmjpa}). It was also pointed out in (\cite{ncqmjpa}) that two certain gauge equivalent representations of NCQM, viz. the Landau and the symmetric gauge representations, arise from two unitarily equivalent representations of $\g$ determined by a fixed value of the triple $(\hbar,\vartheta,\mathcal{B})$.

\section{The algebraic structure associated with the Lie group \texorpdfstring{$\g$}{GNCQM} and the geometry of its coadjoint orbits}\label{sec:algbrc-strctr-review}
The defining group $\g$ of NCQM was first introduced in (\cite{ncqmjmp}) and later in (\cite{ncqmjpa}), the geometry of its coadjoint orbits was studied and subsequently the associated unitary dual $\gd$ was computed. In this section, we shall summarize the relevant results obtained in the two articles.

The group $\g$ is a 7-dimensional real nilpotent Lie group. Its group composition rule is given by (see \cite{ncqmjmp})
\bea\label{grp-law}
\lefteqn{(\theta,\phi,\psi,\bq,\bp)(\theta^{\prime},\phi^{\prime},\psi^{\prime},\bq^{\prime},\bp^{\prime})}\nonumber\\
&&=(\theta+\theta^{\prime}+\frac{\alpha}{2}[\langle\bq,\bp^{\prime}\rangle-\langle\bp,\bq^{\prime}\rangle],\phi+\phi^{\prime}+\frac{\beta}{2}[\bp\wedge\bp^{\prime}],\psi+\psi^{\prime}+\frac{\gamma}{2}[\bq\wedge\bq^{\prime}]\nonumber\\
&&\;\;\;,\bq+\bq^{\prime},\bp+\bp^{\prime}),
\ena
where $\alpha$, $\beta$ and $\gamma$ denote some strictly positive dimensionfull constants associated with the triple central extension. Here, $\bq=(q_1,q_2)$ and $\bp=(p_1,p_2)$. Also, in (\ref{grp-law}), $\langle.,.\rangle$  and $\wedge$ are defined as $\langle\bq,\bp\rangle:=q_{1}p_{1}+q_{2}p_{2}$ and $\bq\wedge\bp:=q_{1}p_{2}-q_{2}p_{1}$, respectively.

It is also important to note that if one denotes by $[q]$ and $[p]$, the dimensions of the position and momentum coordinates, respectively, then, in order to have $\theta$, $\phi$ and $\psi$ to be dimensionless in view of (\ref{grp-law}), one must require that the following holds
\begin{equation}\label{eq:dimension-constnts}
[\alpha]=\left[\frac{1}{pq}\right],\;[\beta]=\left[\frac{1}{p^2}\right],\;\hbox{and}\;[\gamma]=\left[\frac{1}{q^2}\right].
\end{equation}

Let us now quickly recap the geometry of the coadjoint orbits associated with the group $\g$, the detail of which can be found in (\cite{ncqmjpa}). The Lie algebra $\G$ is evidently a 7-dimensional vector space over the reals. Let us choose a set of abstract basis elements of this algebra to be $\{X_1,X_2,..,X_7\}$ so that an arbitrary algebra element $X$ can be written as $X=\sum\limits_{i=1}^{7}x^{i}X_{i}$ with $x^i$'s being the coordinate functions of $X$. Therefore, it is reasonable to choose the coordinate functions of an element $F$ in the dual algebra $\G^*$ to be the set $\{X_1,X_2,..,X_7\}$ with the dual pairing given by $\langle F,X\rangle=\sum\limits_{i=1}^{7}x^{i}X_{i}$. Note that $X_i$'s and hence $X$ are treated as monomials here, not as matrices. Refer to (\cite{ncqmjpa}) to avoid any confusion in this context.

If one denotes a group element having coordinates $p_1$, $p_2$, $q_1$, $q_2$, $\theta$, $\phi$ and $\psi$ by $g(p_1,p_2,q_1,q_2,\theta,\phi,\psi)$, then the coadjoint action K of $\g$ on $\G^*$ reads (p. 5, \cite{ncqmjpa}):
\begin{eqnarray}\label{coad-action-exprssn}
\lefteqn{Kg(p_{1},p_{2},q_{1},q_{2},\theta,\phi,\psi)(X_{1},X_{2},X_{3},X_{4},X_{5},X_{6},X_{7})}\nonumber\\
&&=(X_{1}-\frac{\alpha}{2}q_{1}X_{5}+\frac{\beta}{2}p_{2}X_{6},\;\;X_{2}-\frac{\alpha}{2}q_{2}X_{5}-\frac{\beta}{2}p_{1}X_{6}\nonumber\\
&&\;\;\;\;,X_{3}+\frac{\gamma}{2}q_{2}X_{7}+\frac{\alpha}{2}p_{1}X_{5},\;X_{4}-\frac{\gamma}{2}q_{1}X_{7}+\frac{\alpha}{2}p_{2}X_{5},\;X_{5},\;X_{6},\;X_{7}).
\end{eqnarray}
A somewhat different notation was used for the group coordinates while deriving (\ref{coad-action-exprssn}) in (\cite{ncqmjpa}). But we prefer sticking to the notations of our original group parameters here.

If one denotes the 3-polynomial invariants $X_5$, $X_6$ and $X_7$ by $\rho$, $\sigma$ and $\tau$, respectively, then the underlying coadjoint orbits can be classified based on the values of the triple $(\rho,\sigma,\tau)$ in the following ways:
\begin{itemize}
\item [$ \bullet $] When $\rho\neq 0$, $\sigma\neq 0$ and $\tau\neq 0$ satisfying $\rho^{2}\alpha^{2}-\gamma\beta\sigma\tau\neq 0$, the coadjoint orbits denoted by $\mathcal{O}^{\rho,\sigma,\tau}_{4}$ are $\mathbb{R}^4$, considered as affine 4-spaces.

\item [$ \bullet $] When $\rho\neq 0$, $\sigma\neq 0$ and $\tau\neq 0$ satisfying $\rho^{2}\alpha^{2}-\gamma\beta\sigma\tau=0$, the coadjoint orbits are denoted by $^{\kappa,\delta}\mathcal{O}^{\rho,\zeta}_{2}$. For each ordered pair $(\kappa,\delta)\in\mathbb{R}^{2}$ along with $\rho\neq 0$ and $\zeta\in(-\infty,0)\cup(0,\infty)$ satisfying $\rho=\sigma\zeta=\frac{\gamma\beta\tau}{\zeta\alpha^{2}}$, one obtains an $\mathbb{R}^2$-affine space to be the underlying coadjoint orbit $^{\kappa,\delta}\mathcal{O}^{\rho,\zeta}_{2}$.

\item [$ \bullet $] When $\rho\neq 0$, $\sigma\neq 0$, but $\tau=0$, the coadjoint orbits denoted by $\mathcal{O}^{\rho,\sigma,0}_{4}$ are $\mathbb{R}^4$-affine spaces.

\item [$ \bullet $] When $\rho\neq 0$, $\tau\neq 0$, but $\sigma=0$, the coadjoint orbits denoted by $\mathcal{O}^{\rho,0,\tau}_{4}$ are $\mathbb{R}^4$-affine spaces.

\item [$ \bullet $] When $\rho=0$, $\tau\neq 0$ and $\sigma\neq 0$, the coadjoint orbits denoted by $\mathcal{O}^{0,\sigma,\tau}_{4}$ are also $\mathbb{R}^4$-affine spaces.

\item [$ \bullet $] When $\rho\neq 0$ only but both $\sigma$ and $\tau$ are taken to be identically zero, the coadjoint orbits denoted by $\mathcal{O}^{\rho,0,0}_{4}$ are $\mathbb{R}^4$-affine spaces.

\item [$ \bullet $] When $\rho=\tau=0$ but $\sigma\neq 0$, the underlying coadjoint orbit denoted by $^{c_3,c_4}\mathcal{O}^{0,\sigma,0}_{2}$ is an affine $\mathbb{R}^{2}$-plane. For each fixed ordered pair $(c_3,c_4)$ such a 2-dimensional coadjoint orbit exists.

\item [$ \bullet $] When $\rho=\sigma=0$ but $\tau\neq 0$, the underlying coadjoint orbit denoted by $^{c_1,c_2}\mathcal{O}^{0,0,\tau}_{2}$ is an affine $\mathbb{R}^{2}$-plane. For each fixed ordered pair $(c_1,c_2)$ such a 2-dimensional coadjoint orbit exists.

\item [$ \bullet $] When $\rho=\sigma=\tau=0$, the coadjoint orbits are 0-dimensional points denoted by $^{c_1,c_2,c_3,c_4}\mathcal{O}^{0,0,0}_{0}$. Every quadruple $(c_1,c_2,c_3,c_4)$ gives rise to such an orbit.
\end{itemize}

\section{Classifications of unitary irreducible representations of \texorpdfstring{$\g$}{GNCQM} and those of its Lie algebra \texorpdfstring{$\G$}{ANCQM}}\label{sec:classfctn-rep-grp-algbr}
In this section, we recapitulate the basic results concerning the computations of the equivalence classes of unitary irreducible representations of $\g$ and its Lie algebra $\G$. The details of these computations can be found in (\cite{ncqmjpa}). 

Since, $\g$ is a connected, simply connected nilpotent Lie group, its unitary irreducible representations are in 1-1 correspondence with the underlying coadjoint orbits as corroborated by the method of orbit (see \cite{Kirillovbook}). Therefore, in accordance with the classifications of the coadjoint orbits of $\g$ described in section (\ref{sec:algbrc-strctr-review}), one expects precisely the following nine distinct types of equivalence classes of unitary irreducible representations of $\g$ and its Lie algebra $\G$:

\subsection{Case \texorpdfstring{$\rho\neq 0, \sigma\neq 0, \tau\neq 0$ with $\rho^{2}\alpha^{2}-\gamma\beta\sigma\tau\neq 0.$}{rho sigma tau Nonzero}}\label{subsec:all-nonzero}
The group $\g$ admits a family of unitary irreducible representations $U^{\rho}_{\sigma,\tau}$, defined on $L^{2}(\mathbb{R}^{2},d\br)$, that are associated with its 4-dimensional coadjoint orbits $\mathcal{O}^{\rho,\sigma,\tau}_{4}$. These representations are given by
\bea\label{eq:rep-grp-all-nozer-factr-nozer}
\lefteqn{(U^{\rho}_{\sigma,\tau}(\theta,\phi,\psi,\bq,\bp)f)(\br)}\nonumber\\
    &&=e^{i\rho(\theta+\alpha p_{1}r_{1}+\alpha p_{2}r_{2}+\frac{\alpha}{2}q_{1}p_{1}+\frac{\alpha}{2}q_{2}p_{2})}e^{i\sigma(\phi+\frac{\beta}{2}p_{1}p_{2})}\nonumber\\
    &&\times e^{i\tau(\psi+\gamma q_{2}r_{1}+\frac{\gamma}{2}q_{1}q_{2})}f\left(r_{1}+q_{1},r_{2}+q_{2}+\frac{\sigma\beta}{\rho\alpha}p_{1}\right),
\ena
where $f\in L^{2}(\mathbb{R}^{2},d\br)$.

The irreducible representation of the universal enveloping algebra $\mathcal{U}(\G)$ is realized as self-adjoint differential operators on the smooth vectors of $L^{2}(\mathbb{R}^{2},d\br)$, i.e. the Schwartz space, $\mathcal{S}(\mathbb{R}^{2})$ given by
\begin{equation}\label{algbr-rep-equivalent-nonzero}
    \begin{split}
    &\hat{Q}_{1}=r_{1}+i\vartheta\frac{\partial}{\partial r_{2}},\qquad
    \hat{Q}_{2}=r_{2},\\
    &\hat{P}_{1}=-i\hbar\frac{\partial}{\partial r_{1}},\qquad
    \hat{P}_{2}=-\frac{\bm{\mathcal{B}}}{\hbar}r_{1}-i\hbar\frac{\partial}{\partial r_{2}},
    \end{split}
    \end{equation}
    with the following identification:
    \begin{equation}\label{idntfctn-nonzero-rep}
    \hbar=\frac{1}{\rho\alpha},\;\;\vartheta=-\frac{\sigma\beta}{(\rho\alpha)^{2}}\;\hbox{and}\;\bm{\mathcal{B}}=-\frac{\tau\gamma}{(\rho\alpha)^{2}}.
    \end{equation}
    $B:=\frac{\bm{\mathcal{B}}}{\hbar}$, here, can be interpreted as the constant magnetic field applied normally to the $\hat{Q}_{1}\hat{Q}_{2}$-plane. Using (\ref{idntfctn-nonzero-rep}), one immediately sees that the triple $(\hbar,\vartheta,\mathcal{B})$ determines the coadjoint orbit of $\g$ and hence its unitary irreducible representation in terms of the physically meaningful parameters $\hbar$, $\vartheta$ and $\mathcal{B}$ that we have mentioned in the introduction already.

\subsection{Case \texorpdfstring{$\rho\neq 0, \sigma\neq 0, \tau\neq 0$ with $\rho^{2}\alpha^{2}-\gamma\beta\sigma\tau=0.$}{rho sigma tau Nonzero det Zero}}\label{subsec:all-nonzero-det-zero}
In this case, the unitary irreducible representations defined on $L^{2}(\mathbb{R},dr)$ associated with the 2-dimensional coadjoint orbits $^{\kappa,\delta}\mathcal{O}^{\rho,\zeta}_{2}$ read as
\begin{eqnarray}\label{nonzero-rep-two-dimensional}
    \lefteqn{(U^{\kappa,\delta}_{\rho,\zeta}(\theta,\phi,\psi,q_{1},q_{2},p_{1},p_{2})f)(r)}\nonumber\\
    &&=e^{i\rho\left(\theta+\frac{1}{\zeta}\phi+\frac{\zeta\alpha^{2}}{\gamma\beta}\psi\right)+i\kappa q_{1}+i\delta q_{2}-i\rho\alpha rp_{1}-\frac{i\rho\alpha^{2}\zeta}{\beta}rq_{2}+\frac{i\rho\alpha}{2}(q_{1}p_{1}-q_{2}p_{2})}\nonumber\\
    &&\times e^{i\rho\left(\frac{\alpha^{2}\zeta}{2\beta}q_{1}q_{2}-\frac{\beta}{2\zeta}p_{1}p_{2}\right)}f(r-q_{1}+\frac{\beta}{\alpha\zeta}p_{2}),
\end{eqnarray}
where $f\in L^{2}(\mathbb{R},dr)$.

The relevant representations for the algebra are realized as self-adjoint differential operators acting on smooth vectors of $L^{2}(\mathbb{R},dr)$, i.e. the Schwartz space $\mathcal{S}(\mathbb{R})$ in the following way:
\begin{equation}\label{rep-nonzero-det-zero}
\begin{aligned}
&\hat{Q}_{1}=-r,\quad
\hat{Q}_{2}=i\vartheta\frac{\partial}{\partial r},\\
&\hat{P}_{1}=\hbar\kappa+i\hbar\frac{\partial}{\partial r},\quad
\hat{P}_{2}=\hbar\delta+\frac{\hbar r}{\vartheta},
\end{aligned}
\end{equation}
where we have used the identification given by (\ref{idntfctn-nonzero-rep}).

\subsection{Case \texorpdfstring{$\rho\neq 0, \sigma\neq 0, \tau=0.$}{rho sigma Nonzero}}\label{subsec:rho-sigma-nonzero}
The unitary irreducible representations $U^{\rho}_{\sigma,0}$ associated with the 4-dimensional coadjoint orbits $\mathcal{O}^{\rho,\sigma,0}_{4}$ of $\g$ are given by
\bea\label{eq:rep-grp-rho-sigma-nonzero}
\lefteqn{(U^{\rho}_{\sigma,0}(\theta,\phi,\psi,\bq,\bp)f)(\br)}\nonumber\\
    &&=e^{i\rho(\theta+\alpha p_{1}r_{1}+\alpha p_{2}r_{2}+\frac{\alpha}{2}q_{1}p_{1}+\frac{\alpha}{2}q_{2}p_{2})}e^{i\sigma(\phi+\frac{\beta}{2}p_{1}p_{2})}f\left(r_{1}+q_{1},r_{2}+q_{2}+\frac{\sigma\beta}{\rho\alpha}p_{1}\right),
\ena
where $f\in L^{2}(\mathbb{R}^{2},d\br)$.

The relevant algebra representations realized as self-adjoint differential operators on the Schwartz space $\mathcal{S}(\mathbb{R}^{2})$ then read
\begin{equation}\label{algbr-rep-rho-sigma-nonzero}
    \begin{split}
    &\hat{Q}_{1}=r_{1}+i\vartheta\frac{\partial}{\partial r_{2}},\;\;
    \hat{Q}_{2}=r_{2},\\
    &\hat{P}_{1}=-i\hbar\frac{\partial}{\partial r_{1}},\;\;
    \hat{P}_{2}=-i\hbar\frac{\partial}{\partial r_{2}},
    \end{split}
    \end{equation}
    with the same identification given by (\ref{idntfctn-nonzero-rep}).

\subsection{Case \texorpdfstring{$\rho\neq 0, \sigma= 0, \tau\neq 0.$}{rho tau Nonzero}}\label{subsec:rho-tau-nonzero}
A continuous family of group representations corresponding to the 4-dimensional coadjoint orbits $\mathcal{O}^{\rho,0,\tau}_{4}$ of $\g$ can be obtained using the powerful method of orbit. This family of unitary irreducible representations reads as follows
\bea\label{eq:rep-grp-rho-tau-nozer}
\lefteqn{(U^{\rho}_{0,\tau}(\theta,\phi,\psi,\bq,\bp)f)(\br)}\nonumber\\
    &&=e^{i\rho(\theta+\alpha p_{1}r_{1}+\alpha p_{2}r_{2}+\frac{\alpha}{2}q_{1}p_{1}+\frac{\alpha}{2}q_{2}p_{2})}e^{i\tau(\psi+\gamma q_{2}r_{1}+\frac{\gamma}{2}q_{1}q_{2})}f(\br+\bq),
\ena
where $f\in L^{2}(\mathbb{R}^{2},d\br)$.

The irreducible representations associated with the corresponding algebra can be read off immediately as
\begin{equation}\label{algbr-rep-rho-tau-nonzero}
    \begin{split}
    &\hat{Q}_{1}= r_{1},\;\;
    \hat{Q}_{2}=r_{2},\\
    &\hat{P}_{1}=-i\hbar\frac{\partial}{\partial r_{1}},\;\;
    \hat{P}_{2}=-\frac{\bm{\mathcal{B}}}{\hbar}r_{1}-i\hbar\frac{\partial}{\partial r_{2}},
    \end{split}
    \end{equation}
    where $\hbar$ and $\bm{\mathcal{B}}$ are again given by (\ref{idntfctn-nonzero-rep}).

\subsection{Case \texorpdfstring{$\rho\neq0, \sigma= 0, \tau= 0.$}{rho Nonzero}}\label{subsec:rho-nonzero}
There is a 1-parameter family of unitary irreducible representations of $\g$ that arises from its 4-dimensional coadjoint orbits denoted by $\mathcal{O}^{\rho,0,0}_{4}$. These are precisely the unitary irreducible representations of the 5-dimensional Weyl-Heisenberg group discussed in the introduction (\ref{sec:intro}). The representations, realized on $L^{2}(\mathbb{R}^{2})$, are as follow
\bea\label{eq:rep-grp-rho-nonzer}
\lefteqn{(U^{\rho}_{0,0}(\theta,\phi,\psi,\bq,\bp)f)(\br)}\nonumber\\
    &&=e^{i\rho(\theta+\alpha p_{1}r_{1}+\alpha p_{2}r_{2}+\frac{\alpha}{2}q_{1}p_{1}+\frac{\alpha}{2}q_{2}p_{2})}f(\br+\bq),
\ena
where $f\in L^{2}(\mathbb{R}^{2},d\br)$.

The corresponding irreducible representations of the universal enveloping algebra $\mathcal{U}(\G)$ are given by
\begin{equation}\label{algbr-rep-rho-nonzero}
    \begin{split}
    &\hat{Q}_{1}=r_{1},\;\;
    \hat{Q}_{2}=r_{2},\\
    &\hat{P}_{1}=-i\hbar\frac{\partial}{\partial r_{1}},\;\;
    \hat{P}_{2}=-i\hbar\frac{\partial}{\partial r_{2}},
    \end{split}
    \end{equation}
    
\subsection{Case \texorpdfstring{$\rho=0, \sigma\neq 0, \tau\neq 0.$}{sigma tau Nonzero}}\label{subsec:sigma-tau-nonzero}
The $4$-dimensional coadjoint orbits $\mathcal{O}^{0, \sigma, \tau}_{4}$ of $\g$ gives rise to the following family of its unitary irreducible representations:
\bea\label{eq:rep-grp-sigma-tau-nonzer}
\lefteqn{(U^{0}_{\sigma,\tau}(\theta,\phi,\psi,\bq,\bp)f)(\br)}\nonumber\\
    &&=e^{i\sigma(\phi+\frac{\beta}{2}p_{1}p_{2})}e^{ir_{2}p_{2}}e^{i\tau(\psi+\gamma q_{2}r_{1}+\frac{\gamma}{2}q_{1}q_{2})}f(r_1+q_1,r_{2}+\sigma\beta p_{1}),
\ena
where $f\in L^{2}(\mathbb{R}^{2},d\br)$.

The corresponding irreducible representations of the algebra can be read off as
\begin{equation}\label{algbr-rep-sigma-tau-nonzero}
    \begin{split}
    &\hat{Q}_{1}=i\kappa_{1}\frac{\partial}{\partial r_{2}},\;\;
    \hat{Q}_{2}=r_{2},\\
    &\hat{P}_{1}=-i\frac{\partial}{\partial r_{1}},\;\;
    \hat{P}_{2}=-\kappa_{2}r_{1},
    \end{split}
    \end{equation}
    with $\kappa_{1}=-\sigma\beta$ and $\kappa_{2}=-\tau\gamma$.

The absence of $\rho$ (or $\hbar$ in view of (\ref{idntfctn-nonzero-rep})) in (\ref{algbr-rep-sigma-tau-nonzero}) indicates the fact that we have the noncommutative $q$ and $p$-planes here which don't talk to each other. Hence $q$'s and $p$'s are not be treated as position and momentum coordinates, respectively, rather they are to be considered as dimensionless quantities both in (\ref{eq:rep-grp-sigma-tau-nonzer}) and in (\ref{algbr-rep-sigma-tau-nonzero}).

\subsection{Case \texorpdfstring{$\rho= 0, \sigma= 0, \tau\neq 0.$}{tau Nonzero}}\label{subsec:tau-nonzero}
A continuous family of unitary irreducible representations of $\g$ corresponding to its 2-dimensional coadjoint orbits $^{c_{1}, c_{2}}\mathcal{O}^{0, 0, \tau}_{2}$ for a fixed ordered pair $(c_1,c_2)$ is realized on $L^{2}(\mathbb{R},dr)$ and is given by
\begin{eqnarray}\label{taunonzero-rep-two-dim}
\lefteqn{(U^{c_{1}, c_{2}}_{0, 0, \tau}(\theta,\phi,\psi,\bq,\bp)f)(r)}\nonumber\\
&&=e^{ic_{1}p_{1}+ic_{2}p_{2}}e^{i\tau(\psi-\gamma q_{1}r-\frac{\gamma}{2}q_{1}q_{2})}f(r+q_{2}),
\end{eqnarray}
where $\tau$ is nonzero and $f\in L^{2}(\mathbb{R},dr)$.

The irreducible representations of the universal enveloping algebra $\mathcal{U}(\G)$, realized as self-adjoint differential operators acting on smooth vectors of $L^{2}(\mathbb{R},dr)$, i.e. the Schwartz space $\mathcal{S}(\mathbb{R})$, are given by
\begin{equation}\label{algbr-rep-tau-nonzero}
    \begin{split}
    &\hat{Q}_{1}=c_{1}\mathbb{I},\;\;
    \hat{Q}_{2}=c_{2}\mathbb{I},\\
    &\hat{P}_{1}=\kappa_{2}r,\;\;
    \hat{P}_{2}=-i\frac{\partial}{\partial r},
    \end{split}
    \end{equation}
where $\kappa_{2}=-\tau\gamma$ as in (\ref{algbr-rep-sigma-tau-nonzero}). Physically, this case refers to a noncommutative $p$-plane, i.e. the $\hat{P}_{1}$-$\hat{P}_{2}$-plane.

\subsection{Case \texorpdfstring{$\rho= 0, \sigma\neq 0, \tau= 0.$}{sigma Nonzero}}\label{subsec:sigma-nonzero}
For a fixed ordered pair $(c_3,c_4)$, one can obtain a 1-parameter family of unitary irreducible representations of $\g$ that are associated with the 2-dimensional coadjoint orbits $^{c_{3}, c_{4}}\mathcal{O}^{0, \sigma, 0}_{2}$. These representations, realized on $L^{2}(\mathbb{R},dr)$, are given by
\begin{eqnarray}\label{sigmanonzero-two-dim}
\lefteqn{(U^{c_{3}, c_{4}}_{0, \sigma, 0}(\theta,\phi,\psi,\bq,\bp)f)(r)}\nonumber\\
&&=e^{ic_{3}q_{1}+ic_{4}q_{2}}e^{i\sigma\phi+i rp_{2}+i\frac{\sigma\beta}{2}p_{1}p_{2}}f(r+\sigma\beta p_{1}),
\end{eqnarray}
where $f\in L^{2}(\mathbb{R}, dr)$.

The relevant irreducible representation for the algebra is as follows
\begin{equation}\label{algbr-rep-sigma-nonzero}
    \begin{split}
    &\hat{Q}_{1}=i\kappa_{1}\frac{\partial}{\partial r},\;\;
    \hat{Q}_{2}=r\mathbb{I},\\
    &\hat{P}_{1}=c_{3}\mathbb{I},\;\;
    \hat{P}_{2}=c_{4}\mathbb{I},
    \end{split}
    \end{equation}
where $\mathbb{I}$ is the identity operator on the Schwartz space $\mathcal{S}(\mathbb{R})$ and $\kappa_{1}$ is as given by (\ref{algbr-rep-sigma-tau-nonzero}). Physically, what (\ref{algbr-rep-sigma-nonzero}) represents is just a noncommutative q-plane, i.e. the $\hat{Q}_{1}$-$\hat{Q}_{2}$-plane.

\subsection{Case \texorpdfstring{$\rho= 0, \sigma= 0, \tau= 0.$}{All zero}}\label{subsec:all-zero}
The 1-dimensional representations associated with the 0-dimensional coadjoint orbits of $\g$ due to a fixed quadruple $(c_1,c_2,c_3,c_4)$ are given by
\begin{eqnarray}\label{allzero-rep-zero-dim}
\lefteqn{U^{c_{1}, c_{2}, c_{3}, c_{4}}_{0, 0, 0}(\theta,\phi,\psi,\bq,\bp)}\nonumber\\
&&=e^{ic_{1}p_{1}+ic_{2}p_{2}+ic_{3}q_{1}+ic_{4}q_{2}}.
\end{eqnarray}

The corresponding representation of the algebra is trivial and all the elements of it are mapped to scalar multiple of identity.

\section{On the unitarily equivalent irreducible representations of $\g$ and gauges of NCQM}\label{sec:gauge-sec}

We start this section by noting that there exists a 2-parameter continuous family of equivalent UIRs of the Lie group $\g$, associated with the 4-dimensional generic coadjoint orbits $\mathcal{O}^{\rho,\sigma,\tau}_{4}$ for nonzero $\rho$, $\sigma$ and $\tau$ satisfying $\rho^{2}\alpha^{2}-\tau\gamma\sigma\beta\neq 0$. This family of UIRs of $\g$, in turn, gives rise to the self adjoint representations of $\G$ and motivates the definition of vector potential $\bA$ for a system of NCQM with 2-degrees of freedom. We have the following theorem:

\begin{Theo}\label{2-parmtr-famly-gaug-eqv-rep}
A 2-parameter $(l,m)$ continuous family of unitarily equivalent irreducible representations, associated with the 4-dimensional coadjoint orbit $\mathcal{O}^{\rho,\sigma,\tau}_{4}$ of the connected and simply connected nilpotent Lie group $\g$ due to a fixed nonzero triple $(\rho,\sigma,\tau)$ satisfying $\rho^{2}\alpha^{2}-\tau\gamma\sigma\beta\neq 0$, is given by
\begin{eqnarray}\label{gauge-equiv-irrep}
\lefteqn{(U^{\rho,\sigma,\tau}_{l,m}(\theta,\phi,\psi,\bq,\bp)f)(r_1,r_2)}\nonumber\\
&&=e^{i\rho\theta+i\sigma\phi+i\tau\psi}e^{i\rho\alpha p_{1}r_{1}+i\rho\alpha p_{2}r_{2}+\frac{i\rho^{2}\alpha^{2}\gamma(1-l)}{\tau\gamma\sigma\beta l-\rho^{2}\alpha^{2}}q_{1}r_{2}+il\tau\gamma q_{2}r_{1}+i\left[\frac{\rho\alpha}{2}+\frac{\rho\alpha\tau\gamma\sigma\beta m(1-l)}{\tau\gamma\sigma\beta l-\rho^{2}\alpha^{2}}\right]p_{1}q_{1}}\nonumber\\
&&\times e^{i\left[\frac{\rho\alpha}{2}-\frac{l\tau\gamma\sigma\beta(1-m)}{\rho\alpha}\right]p_{2}q_{2}+i\left(m-\frac{1}{2}\right)\sigma\beta p_{1}p_{2}+i\left[\frac{\tau\gamma}{2}-\frac{\tau\gamma(1-l)(\tau\gamma\sigma\beta l-\tau\gamma\sigma\beta lm-\rho^{2}\alpha^{2})}{\tau\gamma\sigma\beta l-\rho^{2}\alpha^{2}}\right]q_{1}q_{2}}\nonumber\\
&&\scriptstyle\times f\left(r_{1}-\frac{(1-m)\sigma\beta}{\rho\alpha}p_{2}+\frac{\tau\gamma\sigma\beta(l+m-lm)-\rho^{2}\alpha^{2}}{\tau\gamma\sigma\beta l-\rho^{2}\alpha^{2}}q_{1},r_{2}+\frac{m\sigma\beta}{\rho\alpha}p_{1}-\frac{\tau\gamma\sigma\beta l(1-m)-\rho^{2}\alpha^{2}}{\rho^{2}\alpha^{2}}q_{2}\right),
\end{eqnarray}
where $f\in L^{2}(\mathbb{R}^{2},d\br)$. Here, $l\in\mathbb{R}\smallsetminus\left\{\frac{\rho^{2}\alpha^{2}}{\tau\gamma\sigma\beta}\right\}$ and $m\in\mathbb{R}$.
\end{Theo}

\prf{.}
By a rather straightforward but lengthy computation, it can be verified that the operator $U^{\rho,\sigma,\tau}_{l,m}$, defined by its action (\ref{gauge-equiv-irrep}) on $L^{2}(\mathbb{R}^{2},d\br)$, indeed satisfies
\begin{eqnarray}\label{proof-rep}
\lefteqn{(U^{\rho,\sigma,\tau}_{l,m}(\theta,\phi,\psi,\bq,\bp)U^{\rho,\sigma,\tau}_{l,m}(\theta^{\prime},\phi^{\prime},\psi^{\prime},\bq^{\prime},\bp^{\prime})f)(\br)}\nonumber\\
&&=(U^{\rho,\sigma,\tau}_{l,m}((\theta,\phi,\psi,\bq,\bp)(\theta^{\prime},\phi^{\prime},\psi^{\prime},\bq^{\prime},\bp^{\prime}))f)(\br),
\end{eqnarray}
for any $f\in L^{2}(\mathbb{R}^{2},d\br)$. In other words, $U^{\rho,\sigma,\tau}_{l,m}$ defined by (\ref{gauge-equiv-irrep}) is indeed a representation of the Lie group $\g$ obeying the group law (\ref{grp-law}).

Now the adjoint of the representation $U^{\rho,\sigma,\tau}_{l,m}$, given by its action on $L^{2}(\mathbb{R}^{2},d\br)$, can be read off as
\begin{eqnarray}\label{adjoint-gauge-equiv-irrep}
\lefteqn{((U^{\rho,\sigma,\tau}_{l,m})^{*}(\theta,\phi,\psi,\bq,\bp)f)(r_1,r_2)}\nonumber\\
&&=e^{-i\rho\theta-i\sigma\phi-i\tau\psi}e^{-i\rho\alpha p_{1}r_{1}-i\rho\alpha p_{2}r_{2}-\frac{i\rho^{2}\alpha^{2}\tau\gamma(1-l)}{\tau\gamma\sigma\beta l-\rho^{2}\alpha^{2}}q_{1}r_{2}-il\tau\gamma q_{2}r_{1}+i\left[\frac{\rho\alpha}{2}+\frac{\rho\alpha\tau\gamma\sigma\beta m(1-l)}{\tau\gamma\sigma\beta l-\rho^{2}\alpha^{2}}\right]p_{1}q_{1}}\nonumber\\
&&\times e^{i\left[\frac{\rho\alpha}{2}-\frac{l\tau\gamma\sigma\beta(1-m)}{\rho\alpha}\right]p_{2}q_{2}+i\left(m-\frac{1}{2}\right)\sigma\beta p_{1}p_{2}+i\left[\frac{\tau\gamma}{2}-\frac{\tau\gamma(1-l)(\tau\gamma\sigma\beta l-\tau\gamma\sigma\beta lm-\rho^{2}\alpha^{2})}{\tau\gamma\sigma\beta l-\rho^{2}\alpha^{2}}\right]q_{1}q_{2}}\nonumber\\
&&\scriptstyle\times f\left(r_{1}+\frac{(1-m)\sigma\beta}{\rho\alpha}p_{2}-\frac{\tau\gamma\sigma\beta(l+m-lm)-\alpha^{2}}{\tau\gamma\sigma\beta l-\rho^{2}\alpha^{2}}q_{1},r_{2}-\frac{m\sigma\beta}{\rho\alpha}p_{1}+\frac{\tau\gamma\sigma\beta l(1-m)-\rho^{2}\alpha^{2}}{\rho^{2}\alpha^{2}}q_{2}\right).
\end{eqnarray}

By direct substitution, one can now immediately check that the following equality holds
\begin{equation}\label{unitarity-check}
((U^{\rho,\sigma,\tau}_{l,m})^{*}U^{\rho,\sigma,\tau}_{l,m}f)(\br)=(U^{\rho,\sigma,\tau}_{l,m}(U^{\rho,\sigma,\tau}_{l,m})^{*}f)(\br)=f(\br),
\end{equation}
for any $f\in L^{2}(\mathbb{R}^{2},d\br)$. In other words, the representations $U^{\rho,\sigma,\tau}_{l,m}$ of $\g$ given by (\ref{gauge-equiv-irrep}), are indeed unitary. What just remains to be proven that they are also irreducible. 

The corresponding irreducible representation of the Lie algebra $\G$ by self-adjoint operators on the smooth vectors of $L^{2}(\mathbb{R}^{2},d\br)$, i.e. the Schwartz space $\mathcal{S}(\mathbb{R}^{2})$, is given below by (\ref{gauge-equiv-reps-algbr}). The nonvanishing commutation relations between the underlying self-adjoint operators read
\begin{equation}\label{non-vanishing-commut-rel-algbr-bas}
\begin{split}
&[\hat{Q}_{1},\hat{P}_{1}]=[\hat{Q}_{2},\hat{P}_{2}]=\frac{i}{\rho\alpha}\mathbb{I},\\
&[\hat{Q}_{1},\hat{Q}_{2}]=-\frac{i\sigma\beta}{\rho^{2}\alpha^{2}}\mathbb{I},\\
&[\hat{P}_{1},\hat{P}_{2}]=-\frac{i\tau\gamma}{\rho^{2}\alpha^{2}}\mathbb{I},
\end{split}
\end{equation}
with $\mathbb{I}$ being the identity operator on $L^{2}(\mathbb{R}^{2},d\br)$. It can now be easily verified that (\ref{non-vanishing-commut-rel-algbr-bas}), indeed, satisfies (\ref{eq:NCCR}) using the identification (\ref{idntfctn-nonzero-rep}). Now, the condition $\rho^{2}\alpha^{2}-\tau\gamma\sigma\beta\neq 0$ attributes irreducibility to the representation (\ref{non-vanishing-commut-rel-algbr-bas}).  It means that there exists no proper subspace of $\mathcal{S}(\mathbb{R}^{2})$, the smooth vectors of which will stay invariant under the action of the self-adjoint operators given by (\ref{gauge-equiv-reps-algbr}). Indeed, when $\rho^{2}\alpha^{2}-\tau\gamma\sigma\beta\rightarrow 0$, one can check by straightforward manipulation that $\hat{P}_{1}^{l,m}+\frac{\tau\gamma}{\rho\alpha}\hat{Q}_{2}^{m}\rightarrow 0$ holds; in other words, $\hat{P}^{l,m}_{1}$ and $\hat{Q}_{2}^{m}$ become proportional to each other in this limiting case, turning $\mathcal{S}(\mathbb{R}^{2})$ too big to represent $\G$ irreducibly on it using (\ref{gauge-equiv-reps-algbr}).

Hence using the fact that $\g$ is a connected, simply connected  Lie group, one can conclude that the pertinent unitary representations (\ref{gauge-equiv-irrep}) of $\g$ are also irreducible. Indeed, by the orbit method, one can compute an irreducible unitary representation ( \ref{eq:rep-grp-all-nozer-factr-nozer}) of $\g$. This irreducible representation belongs to the 2-parameter family (\ref{gauge-equiv-irrep}) due to $l=m=1$. Therefore, all other members of the underlying family have to be unitarily equivalent to each other. These unitarily equivalent irreducible representations of $\g$ are actually gauge equivalent in the sense to be discussed following definition \ref{Vect-pot-def}.

\qed

A continuous family of unitarily equivalent irreducible representations of the Lie algebra $\G$ parameterized by a real  ordered pair $(l,m)$ can be realized as self-adjoint differential operators acting on the smooth vectors of the Schwartz space $\mathcal{S}(\mathbb{R}^{2})$ in the following way:
\begin{equation}\label{gauge-equiv-reps-algbr}
\begin{split}
&\hat{Q}^{m}_{1}=r_{1}-m\frac{i\sigma\beta}{\rho^{2}\alpha^{2}}\frac{\partial}{\partial r_{2}},\\
&\hat{Q}^{m}_{2}=r_{2}+(1-m)\frac{i\sigma\beta}{\rho^{2}\alpha^{2}}\frac{\partial}{\partial r_{1}},\\
&\hat{P}^{l,m}_{1}=\frac{\tau\gamma\rho\alpha(1-l)}{\tau\gamma\sigma\beta l-\rho^{2}\alpha^{2}}r_{2}-\frac{i}{\rho\alpha}\left[\frac{\tau\gamma\sigma\beta(l+m-lm)-\rho^{2}\alpha^{2}}{\tau\gamma\sigma\beta l-\rho^{2}\alpha^{2}}\right]\frac{\partial}{\partial r_{1}},\\
&\hat{P}^{l,m}_{2}=\frac{l\tau\gamma}{\rho\alpha}r_{1}+i\left[\frac{\tau\gamma\sigma\beta l(1-m)-\rho^{2}\alpha^{2}}{\rho^{3}\alpha^{3}}\right]\frac{\partial}{\partial r_{2}}.
\end{split}
\end{equation}

Rearrange the terms of the last two equations of (\ref{gauge-equiv-reps-algbr}) to obtain
\begin{equation}\label{rearrange-gauge-equiv-reps-algbr}
\begin{split}
&\hat{Q}^{m}_{1}=r_{1}-m\frac{i\sigma\beta}{\rho^{2}\alpha^{2}}\frac{\partial}{\partial r_{2}},\\
&\hat{Q}^{m}_{2}=r_{2}+(1-m)\frac{i\sigma\beta}{\rho^{2}\alpha^{2}}\frac{\partial}{\partial r_{1}},\\
&\hat{P}^{l,m}_{1}=\frac{\tau\gamma\rho\alpha(1-l)}{\tau\gamma\sigma\beta l-\rho^{2}\alpha^{2}}\hat{Q}^{m}_{2}-\frac{i}{\rho\alpha}\left[\frac{\tau\gamma\sigma\beta(1-l)}{\tau\gamma\sigma\beta l-\rho^{2}\alpha^{2}}+1\right]\frac{\partial}{\partial r_{1}},\\
&\hat{P}^{l,m}_{2}=\frac{l\tau\gamma}{\rho\alpha}\hat{Q}^{m}_{1}-\frac{i}{\rho\alpha}\left(1-\frac{l\tau\gamma\sigma\beta}{\rho^{2}\alpha^{2}}\right)\frac{\partial}{\partial r_{2}}.
\end{split}
\end{equation}
Now, (\ref{rearrange-gauge-equiv-reps-algbr}) motivates us to define a 2-parameter family of vector potentials $\bA_{l,m}$ associated with the NCQM gauges for a system with 2-degrees of freedom.

\begin{defi}\label{Vect-pot-def}
Associated with the UIRs (\ref{gauge-equiv-irrep}) of $\g$, one can define the 2-parameter family of vector potentials $\bA^{l,m}\equiv\left(-\frac{\tau\gamma\rho\alpha(1-l)}{\tau\gamma\sigma\beta l-\rho^{2}\alpha^{2}}\hat{Q}^{m}_{2},-\frac{l\tau\gamma}{\rho\alpha}\hat{Q}^{m}_{1}\right)$ for a fixed nonzero triple $(\rho,\sigma,\tau)$ satisfying $\rho^{2}\alpha^{2}-\tau\gamma\sigma\beta\neq 0$, with $\hat{Q}^{m}_{i}$'s as given in (\ref{rearrange-gauge-equiv-reps-algbr}), to be noncommutative vector potentials determining continuous family of NCQM gauges for $l\in\mathbb{R}\smallsetminus\left\{\frac{\rho^{2}\alpha^{2}}{\tau\gamma\sigma\beta}\right\}$ and $m\in\mathbb{R}$. While writing the vector potential $\bA^{l,m}$, its dependence on $\rho$, $\sigma$ and $\tau$ is suppressed due to notational convenience. 
\end{defi}

\begin{remark}
A few remarks on theorem \ref{2-parmtr-famly-gaug-eqv-rep} and about some consequences of definition \ref{Vect-pot-def}   are in order. First of all, the well-known Landau gauge and the symmetric gauge of NCQM belong to the family $\bA^{l,m}$ for $l=1,m=0$ and $l=\frac{\rho\alpha(\rho\alpha-\sqrt{\rho^{2}\alpha^{2}-\tau\gamma\sigma\beta})}{\tau\gamma\sigma\beta}:=l_{s},m=\frac{1}{2}$, respectively. Secondly,  if one denotes the components of $\bA^{l,m}$ with $A^{l,m}_{i}, i=1,2$, then for Landau gauge potential $\bA^{1,0}$, one finds that $\partial_{1}A^{1,0}_{2}-\partial_{2}A^{1,0}_{1}=B$ holds where the applied constant vertical magnetic field is given by $B=-\frac{\tau\gamma}{\rho\alpha}$ (see (3.6) of \cite{ncqmjpa}). Additionally, for symmetric gauge potential $\bA^{l_{s},\frac{1}{2}}$, one verifies that $\partial_{1}A^{l_{s},\frac{1}{2}}_{2}-\partial_{2}A^{l_{s},\frac{1}{2}}_{1}=\bar{B}$ holds with the spatial noncommutativity dependent quantity $\bar{B}$ being given by $\bar{B}=\frac{2\hbar}{\vartheta}\left(\sqrt{1-\frac{B\vartheta}{\hbar}}-1\right)$ (see (3.6) of \cite{ncqmjpa} to obtain $\hbar$, $\vartheta$ and $B$ in terms of $\rho$, $\sigma$ and $\tau$). These results of Landau and symmetric gauges are in exact agreement with what Delduc et al. found in \cite{delducetalt} (see p.14-15). Thirdly, the irreducible representation of $\g$ and that of $\G$ associated with Landau gauge, can be obtained from (\ref{gauge-equiv-irrep}) and (\ref{gauge-equiv-reps-algbr}), respectively, by substituting $l=1$ and $m=0$. The corresponding irreducible representations associated with the symmetric gauge (p.19, \cite{ncqmjpa}) can be deduced by choosing $l=\frac{\rho\alpha(\rho\alpha-\sqrt{\rho^{2}\alpha^{2}-\tau\gamma\sigma\beta})}{\tau\gamma\sigma\beta}$ and $m=\frac{1}{2}$ in (\ref{gauge-equiv-irrep}) and (\ref{gauge-equiv-reps-algbr}). Note that the value of the parameter $l$ here that yields the symmetric gauge, i.e. $l_{s}$, is just a dimensionless real number which can be verified using the dimensions of $\alpha$, $\beta$ and $\gamma$ from (\ref{eq:dimension-constnts}), directly. 

It is also noteworthy that the gauge potential defined in (\ref{Vect-pot-def}) does not in general satisfy $\partial_{1}A^{l,m}_{2}-\partial_{2}A^{l,m}_{1}=-\frac{\tau\gamma}{\rho\alpha}=B$ for all values of $l$ and $m$ with $A^{l,m}_{i}$'s denoting the components of the underlying vector potential. In what follows, we make $\rho$, $\sigma$ and $\tau$ dependence of the vector potential explicit by denoting it with $\bA^{\rho,\sigma,\tau}_{l,m}$ and its components by $A^{\rho,\sigma,\tau}_{l,m,i}$ with $i=1,2$. One then expects to recover the usual Landau mechanics with the spatial noncommutativity approaching zero, i.e. $\lim_{\sigma\to 0}(\partial_{1}A^{\rho,\sigma,\tau}_{l,m,2}-\partial_{2}A^{\rho,\sigma,\tau}_{l,m,1})=B$, where, the applied vertical constant magnetic field is, as before, given by $B=-\frac{\tau\gamma}{\rho\alpha}$.
\end{remark}

We stress the fact before closing this section that the NCQM gauges considered here only concern the family of 4-dimensional coadjoint orbits $\mathcal{O}^{\rho,\sigma,\tau}_{4}$ for nonzero $\rho$, $\sigma$ and $\tau$ satisfying $\rho^{2}\alpha^{2}-\tau\gamma\sigma\beta\neq 0$. We propose to study NCQM gauges for the other coadjoint orbits of $\g$ in a future publication.

\section{Noncommutative 4-tori from the unitary dual \texorpdfstring{$\gd$}{DGNCQM} of the defining group \texorpdfstring{$\g$}{GNCQM} of NCQM}\label{sec:NC-tori-NCQM}
This section is devoted to the study of noncommutative 4-tori that can be constructed out of the equivalence classes of unitary irreducible representations of the defining group $\g$ of NCQM described in section (\ref{sec:classfctn-rep-grp-algbr}).

A noncommutative n-tori or the algebra of smooth functions on noncommutative n-tori to be more precise, abbreviated as NC n-tori in the sequel and denoted by $\mathcal{A}^{n}_{\theta}=C^{\infty}(\mathbb{T}^{n}_{\theta})$,  is a family of noncommutative C* algebras generated by n unitaries subject to the following defining relations:
\be\label{def:NC-tori}
U_{k}U_{j}=e^{2\pi i\theta_{jk}}U_{j}U_{k},
\en
where $j,k=1,2,..,n$ and $\theta=[\theta_{jk}]$ is a skew-symmetric $n\times n$ matrix. When $\theta$ is the zero matrix, the C* algebra generated by $U_j$'s is a commutative one and can be identified with the continuous functions on the n-torus. A noncommutative n-torus $\mathcal{A}^{n}_{\theta}$ for some skew-symmetric $n\times n$ matrix is called {\em irrational} if not all the entries of $\theta$ are rational. 

We are particularly interested in the case $n=4$ with 4 generators $U_1$, $U_2$, $U_3$ and $U_4$, satisfying the relations given by (\ref{def:NC-tori}). We will compute these generators of NC 4-tori from the unitary dual $\gd$ of the defining group $\g$ of NCQM, introduced in section \ref{sec:classfctn-rep-grp-algbr} and hence construct the skew-symmetric $4\times 4$ matrix $\theta$ due to different levels of underlying noncommutativity (9 distinct types of equivalence classes outlined in section \ref{sec:classfctn-rep-grp-algbr}).

Let us refer back to (\ref{eq:rep-grp-all-nozer-factr-nozer}) and compute the following 4-one parameter groups of unitary operators acting on $L^{2}(\mathbb{R}^{2},d\br)$:
\begin{equation}\label{1-prmtr-grp-all-nonzr-factr-nonzer}
\begin{split}
&(U(q_1)f)(\br)=f(r_1+q_1,r_2)\\
&(U(q_2)f)(\br)=e^{i\tau\gamma q_{2}r_{1}}f(r_1,r_{2}+q_{2})\\
&(U(p_1)f)(\br)=e^{i\rho\alpha p_{1}r_{1}}f\left(r_1,r_{2}+\frac{\sigma\beta}{\rho\alpha}p_1\right)\\
&(U(p_2)f)(\br)=e^{i\rho\alpha p_{2}r_{2}}f(\br),
\end{split}
\end{equation}
obeying the following set of Weyl commutation relations:
\begin{equation}\label{Weyl-commutnt-first-case}
\begin{split}
&U(q_1)U(p_1)=e^{i\rho\alpha q_{1}p_{1}}U(p_1)U(q_1)\\
&U(q_2)U(p_2)=e^{i\rho\alpha q_{2}p_{2}}U(p_2)U(q_2)\\
&U(q_1)U(q_2)=e^{i\tau\gamma q_{1}q_{2}}U(q_2)U(q_1)\\
&U(p_1)U(p_2)=e^{i\sigma\beta p_{1}p_{2}}U(p_2)U(p_1)\\
&U(q_1)U(p_2)=U(p_2)U(q_1)\\
&U(q_2)U(p_1)=U(p_1)U(q_2).
\end{split}
\end{equation}
Let us now suppress the group parameters $q_1$, $q_2$, $p_1$ and $p_2$ by taking 
\begin{equation}\label{constitent-dimension}
\alpha q_{1}p_{1}=\alpha q_{2}p_{2}=2\pi=\gamma q_{1}q_{2}=\beta p_{1}p_{2}
\end{equation}
in (\ref{Weyl-commutnt-first-case}) and denote the unitary operators $U(q_1)$, $U(q_2)$, $U(p_1)$ and $U(p_2)$ by $U_1$, $U_2$, $U_3$ and $U_4$, respectively. Note that (\ref{constitent-dimension}) agrees with the dimensions of the constants $\alpha$, $\beta$ and $\gamma$ as specified in (\ref{eq:dimension-constnts}). The relations (\ref{Weyl-commutnt-first-case}) between the unitary operators $U_1$, $U_2$, $U_3$ and $U_4$ can then be recast as
\begin{equation}\label{recast-Weyl-commutnt-first-case}
\begin{split}
&U_{1}U_{3}=e^{2\pi i\rho}U_{3}U_{1}\\
&U_{2}U_{4}=e^{2\pi i\rho}U_{4}U_{2}\\
&U_{1}U_{2}=e^{2\pi i\tau}U_{2}U_{1}\\
&U_{3}U_{4}=e^{2\pi i\sigma}U_{4}U_{3}\\
&U_{1}U_{4}=U_{4}U_{1}\\
&U_{2}U_{3}=U_{3}U_{2}.
\end{split}
\end{equation}
Comparison of (\ref{recast-Weyl-commutnt-first-case}) with (\ref{def:NC-tori}) yields the skew-symmetric matrix $\theta(\rho,\sigma,\tau)$ with each of $\rho$, $\sigma$ and $\tau$ being nonzero satisfying the inequality $\rho^{2}-\sigma\tau\neq 0$ ( note that this is synonymous with $\rho^{2}\alpha^{2}-\gamma\beta\sigma\tau\neq 0$ as $\alpha^{2}=\gamma\beta$, being a consequence of (\ref{constitent-dimension}), holds).
\begin{equation}\label{skew-symm-matrx-all-nonzer-fact-nonzer}
\theta(\rho,\sigma,\tau)=\begin{bmatrix}0&-\tau&-\rho&0\\\tau&0&0&-\rho\\\rho&0&0&-\sigma\\0&\rho&\sigma&0\end{bmatrix}.
\end{equation}
We denote the family of C* algebras, generated by the unitaries $U_1$, $U_2$, $U_3$ and $U_4$ obeying the relations (\ref{recast-Weyl-commutnt-first-case}), with $\mathcal{A}^{4}_{\theta(\rho,\sigma,\tau)}$ where $\theta(\rho,\sigma,\tau)$ is the skew-symmetric $4\times 4$ matrix given by (\ref{skew-symm-matrx-all-nonzer-fact-nonzer}). Each member of the family $\mathcal{A}^{4}_{\theta(\rho,\sigma,\tau)}$ of C* algebras is associated with one and only 4-dimensional coadjoint orbit $\mathcal{O}^{\rho,\sigma,\tau}_{4}$ of $\g$ enumerated in section \ref{sec:algbrc-strctr-review}.

We now turn our attention to the case when each of $\rho$, $\sigma$ and $\tau$ is nonzero satisfying $\rho^{2}\alpha^{2}-\gamma\beta\sigma\tau=0$. This case deals with the degenerate UIRs (see (\ref{nonzero-rep-two-dimensional})) of $\g$ represented on $L^{2}(\mathbb{R}, dr)$. It defines an elliptic cone-shaped surface with two perpendicular lines deleted in $\mathbb{R}^{3}$ (see the illustration on p.6 of \cite{ncqmjpa}). A point on such a surface lies on the line 
\begin{equation}\label{eq:straight-line}
\rho=\sigma\zeta=\frac{\gamma\beta\tau}{\zeta\alpha^{2}},
\end{equation}
going through the origin and can be uniquely specified by an ordered pair $(\rho,\zeta)$ with $\zeta\in(-\infty,0)\cup(0,\infty)$. For each such point, one can obtain a quadruple of one parameter groups of unitary operators acting on $L^{2}(\mathbb{R},dr)$ from a family of unitary irreducible representations (\ref{nonzero-rep-two-dimensional}) of $\g$ in the following way:
\begin{equation}\label{uniraty-oprtrs-all-nonzer-factr-nonzer}
\begin{split}
&(U(q_1)f)(r)=e^{i\kappa q_{1}}f(r-q_{1})\\
&(U(q_2)f)(r)=e^{i\left(\delta-\frac{\rho\alpha^{2}\zeta}{\beta}r\right)q_{2}}f(r)\\
&(U(p_1)f)(r)=e^{-i\rho\alpha rp_{1}}f(r)\\
&(U(p_2)f)(r)=f(r+\frac{\beta}{\alpha\zeta}p_{2}).
\end{split}
\end{equation}
The one parameter groups of unitary operators given above in (\ref{uniraty-oprtrs-all-nonzer-factr-nonzer}) with their respective actions on $L^{2}(\mathbb{R},dr)$ can also be seen to satisfy the Weyl commutation relations listed in (\ref{Weyl-commutnt-first-case}) if one just expresses $\sigma$ and $\tau$ in terms of $\rho$ and $\zeta$ using the straight line equation (\ref{eq:straight-line}). Suppressing the group parameters (see \ref{constitent-dimension}) in the resulting Weyl commutation relations, one can then obtain 4-unitaries that satisfy the following relations:
\begin{equation}\label{recast-Weyl-commutnt-second-case}
\begin{split}
&U_{1}U_{3}=e^{2\pi i\rho}U_{3}U_{1}\\
&U_{2}U_{4}=e^{2\pi i\rho}U_{4}U_{2}\\
&U_{1}U_{2}=e^{2\pi i\rho\zeta}U_{2}U_{1}\\
&U_{3}U_{4}=e^{\frac{2\pi i\rho}{\zeta}}U_{4}U_{3}\\
&U_{1}U_{4}=U_{4}U_{1}\\
&U_{2}U_{3}=U_{3}U_{2}.
\end{split}
\end{equation}
We now form the 2-parameter family of C* algebras generated by the above 4 unitaries. This family is denoted by $\mathcal{A}^{4}_{\theta(\rho,\zeta)}$ where $\theta(\rho,\zeta)$, being a 2-parameter skew-symmetric $4\times 4$ matrix, can be read off immediately by combining (\ref{recast-Weyl-commutnt-second-case}) with (\ref{def:NC-tori}, $n=4$):
\begin{equation}\label{skew-symm-matrx-all-nonzer-fact-zero}
\theta(\rho,\zeta)=\begin{bmatrix}0&-\rho\zeta&-\rho&0\\\rho\zeta&0&0&-\rho\\\rho&0&0&-\frac{\rho}{\zeta}\\0&\rho&\frac{\rho}{\zeta}&0\end{bmatrix}.
\end{equation}
Note the absence of $\kappa$ and $\delta$ in the Weyl commutation relations (\ref{recast-Weyl-commutnt-second-case}) and hence in the entries of the skew-symmetric matrix given by (\ref{skew-symm-matrx-all-nonzer-fact-zero}). It just reflects the fact that each member of the family $\mathcal{A}^{4}_{\theta(\rho,\zeta)}$ of C* algebras due to a fixed ordered pair $(\rho,\zeta)$ can be associated with a whole slew of 2-dimensional coadjoint orbits $^{\kappa,\delta}\mathcal{O}^{\rho,\zeta}_{2}$ of $\g$ (see section \ref{sec:algbrc-strctr-review}) by varying $\kappa$ and $\delta$ independently.

Following exactly the same steps as adopted in the previous two cases, it is now easy to find 4 unitaries $U_1$, $U_2$, $U_3$ and $U_4$ from the 2-parameter family of irreducible unitary representations (\ref{eq:rep-grp-rho-sigma-nonzero}) of $\g$ that will satisfy the following relations:
\begin{equation}\label{recast-Weyl-commutnt-third-case}
\begin{split}
&U_{1}U_{3}=e^{2\pi i\rho}U_{3}U_{1}\\
&U_{2}U_{4}=e^{2\pi i\rho}U_{4}U_{2}\\
&U_{1}U_{2}=U_{2}U_{1}\\
&U_{3}U_{4}=e^{2\pi i\sigma}U_{4}U_{3}\\
&U_{1}U_{4}=U_{4}U_{1}\\
&U_{2}U_{3}=U_{3}U_{2}.
\end{split}
\end{equation}
The family of C* algebras generated by these unitaries will be denoted by $\mathcal{A}^{4}_{\theta(\rho,\sigma)}$ where $\theta(\rho,\sigma)$ is a skew-symmetric $4\times 4$ matrix given by
\begin{equation}\label{skew-symm-matrx-rho-sigma-nonzer}
\theta(\rho,\sigma)=\begin{bmatrix}0&0&-\rho&0\\0&0&0&-\rho\\\rho&0&0&-\sigma\\0&\rho&\sigma&0\end{bmatrix}.
\end{equation}
Each member of the family $\mathcal{A}^{4}_{\theta(\rho,\sigma)}$ due to fixed nonzero $\rho$ and $\sigma$ is associated with a 4-dimensional coadjoint orbit $\mathcal{O}^{\rho,\sigma,0}_{4}$ of $\g$.

The case $\rho\neq 0$, $\tau\neq 0$, but $\sigma=0$ is similar to the previous situation. Here also, 4 unitaries can be obtained using the irreducible unitary representations (\ref{eq:rep-grp-rho-tau-nozer})of $\g$ that obey the following relations:
\begin{equation}\label{recast-Weyl-commutnt-fourth-case}
\begin{split}
&U_{1}U_{3}=e^{2\pi i\rho}U_{3}U_{1}\\
&U_{2}U_{4}=e^{2\pi i\rho}U_{4}U_{2}\\
&U_{1}U_{2}=e^{2\pi i\tau}U_{2}U_{1}\\
&U_{3}U_{4}=U_{4}U_{3}\\
&U_{1}U_{4}=U_{4}U_{1}\\
&U_{2}U_{3}=U_{3}U_{2}.
\end{split}
\end{equation}
The family of C* algebras generated by the above unitaries will be denoted by $\mathcal{A}^{4}_{\theta(\rho,\tau)}$ where the skew-symmetric matrix $\theta(\rho,\tau)$ is given by
\begin{equation}\label{skew-symm-matrx-rho-tau-nonzer}
\theta(\rho,\tau)=\begin{bmatrix}0&-\tau&-\rho&0\\\tau&0&0&-\rho\\\rho&0&0&0\\0&\rho&0&0\end{bmatrix}.
\end{equation} 
A C* algebra, belonging to the family $\mathcal{A}^{4}_{\theta(\rho,\tau)}$ due to fixed nonzero $\rho$ and $\tau$, can be assigned to a 4-dimensional coadjoint orbit $\mathcal{O}^{\rho,0,\tau}_{4}$ of $\g$.

The one parameter family of irreducible unitary representations (\ref{eq:rep-grp-rho-nonzer}) of $\g$ can be exploited to compute 4 unitaries that satisfy the relations (\ref{def:NC-tori}) with $n=4$ where the skew-symmetric $4\times 4$ matrix is given by
\begin{equation}\label{skew-symm-matrx-rho-nonzer}
\theta(\rho)=\begin{bmatrix}0&0&-\rho&0\\0&0&0&-\rho\\\rho&0&0&0\\0&\rho&0&0\end{bmatrix}.
\end{equation}
The one parameter family of C* algebras, generated by the above unitaries, is denoted by $\mathcal{A}^{4}_{\theta(\rho)}$ where $\theta(\rho)$ is the $4\times 4$ matrix as given above in (\ref{skew-symm-matrx-rho-nonzer}). Each member of such a family due to a fixed $\rho$ can be associated with a 4-dimensional coadjoint orbit $\mathcal{O}^{\rho,0,0}_{4}$ of $\g$. Note that the irreducible unitary representations (\ref{eq:rep-grp-rho-nonzer}) of $\g$ are indeed those of the Weyl-Heisenberg group in 2-dimensions. Therefore, the one parameter family of  noncommutative C* algebras $\mathcal{A}^{4}_{\theta(\rho)}$ is precisely the one associated with standard nonrelativistic quantum mechanics in 2-dimensions.

For $\rho=0$ and nonzero $\sigma$ and $\tau$, the 4 unitary operators defined on $L^{2}(\mathbb{R}^{2},d\br)$ can be computed using the irreducible unitary representations (\ref{eq:rep-grp-sigma-tau-nonzer}) of $\g$ that obey the relations given by (\ref{def:NC-tori}) with $n=4$. The underlying skew-symmetric $4\times 4$ matrix is given by
\begin{equation}\label{skew-symm-matrx-rho-zer-others-nonzer}
\theta(\sigma,\tau)=\begin{bmatrix}0&-\tau&0&0\\\tau&0&0&0\\0&0&0&-\sigma\\0&0&\sigma&0\end{bmatrix}.
\end{equation}
We denote by $\mathcal{A}^{4}_{\theta(\sigma,\tau)}$ the family of C* algebras generated by such unitaries. The members of this family are in 1-1 correspondence with the 4-dimensional coadjoint orbits $\mathcal{O}^{0,\sigma,\tau}_{4}$ of $\g$.

There are two other distinct families of 2-dimensional coadjoint orbits of $\g$ that we denoted by $^{c_{1}, c_{2}}\mathcal{O}^{0, 0, \tau}_{2}$ and $^{c_{3}, c_{4}}\mathcal{O}^{0, \sigma, 0}_{2}$ in section \ref{sec:algbrc-strctr-review}. The unitary irreducible representations associated with these two families of coadjoint orbits are given by (\ref{taunonzero-rep-two-dim}) and (\ref{sigmanonzero-two-dim}), respectively. Using these representations of $\g$, one can compute 4 unitaries that will obey the relations (\ref{def:NC-tori}) with $n=4$. We shall denote the family of C* algebras generated by such unitaries for the case $\tau\neq 0$ and $\sigma\neq 0$ by $\mathcal{A}^{4}_{\theta(\tau)}$ and $\mathcal{A}^{4}_{\theta(\sigma)}$, respectively. Here the skew symmetric $4\times 4$ matrices $\theta(\tau)$ and $\theta(\sigma)$ are given by
\begin{equation}\label{skew-symm-matrx-tau-nonzer--sigma-nonzer}
\theta(\tau)=\begin{bmatrix}0&-\tau&0&0\\\tau&0&0&0\\0&0&0&0\\0&0&0&0\end{bmatrix}\quad\hbox{and}\quad\theta(\sigma)=\begin{bmatrix}0&0&0&0\\0&0&0&0\\0&0&0&-\sigma\\0&0&\sigma&0\end{bmatrix},
\end{equation}
respectively. It is easy to check that the Weyl commutation relations among the unitaries in both the above cases do not involve the coadjoint orbit parameters $c_i$'s for $i=1,2,..,4$. It indicates the fact that the multitude of coadjoint orbits $^{c_{1}, c_{2}}\mathcal{O}^{0, 0, \tau}_{2}$ and $^{c_{3}, c_{4}}\mathcal{O}^{0, \sigma, 0}_{2}$ of $\g$ give rise to the C* algebras $\mathcal{A}^{4}_{\theta(\tau)}$ and $\mathcal{A}^{4}_{\theta(\sigma)}$ due to nonzero fixed values of $\tau$ and $\sigma$, respectively.

The final case to be considered is the one when each of $\rho$, $\sigma$ and $\tau$ is zero. The respective 1-dimensional unitary irreducible representations of $\g$ are given by (\ref{allzero-rep-zero-dim}). The 4 unitaries can easily be seen to commute with each other and hence the skew-symmetric matrix satisfying (\ref{def:NC-tori}) for $n=4$ in this case is just the 0-matrix. We shall denote the commutative C* algebra generated by such unitaries by $\mathcal{A}^{4}_{0}$. This commutative C* algebra can be identified with the algebra of smooth functions on the ordinary 4-torus. 

In view of the above discussions, we therefore have the following theorem:

\begin{Theo}\label{Theo:NC-tori}
The noncommutative 4-tori associated with the noncommutative quantum mechanics in 2-dimensions is a family of C* algebras $\mathcal{A}^{4}_{\theta}$ generated by 4 unitaries subject to the relations (\ref{def:NC-tori}) with $n=4$. 

Let $\mathbb{S}_{\rho,\zeta}=\{(\rho,\sigma,\tau)\in\mathbb{R}^{3}\vert\;\rho\neq 0,\sigma\neq 0,\tau\neq 0\;\hbox{and}\;\rho^{2}-\sigma\tau=0\}$ with the use of the subscript $\rho$ and $\zeta$ here being justified by the fact that any point on the surface defined by $\rho^{2}-\sigma\tau=0$ with nonzero $\rho$, $\sigma$ and $\tau$ lies on the straight line given by $\rho=\sigma\zeta=\frac{\tau}{\zeta}$ for $\zeta\in(-\infty,0)\cup(0,\infty)$.

Here the skew-symmetric $4\times 4$ matrix $\theta$ is given by
\begin{equation}
\begin{split}
&\theta=\begin{bmatrix}0&-\tau&-\rho&0\\\tau&0&0&-\rho\\\rho&0&0&-\sigma\\0&\rho&\sigma&0\end{bmatrix}\quad\hbox{when}\quad(\rho,\sigma,\tau)\in \mathbb{R}^{3}\setminus\mathbb{S}_{\rho,\zeta},\\
&\theta=\begin{bmatrix}0&-\rho\zeta&-\rho&0\\-\rho\zeta&0&0&-\rho\\\rho&0&0&-\frac{\rho}{\zeta}\\0&\rho&\frac{\rho}{\zeta}&0\end{bmatrix}\quad\hbox{when}\quad(\rho,\sigma,\tau)\in\mathbb{S}_{\rho,\zeta}.
\end{split}
\end{equation}

\end{Theo}

\section{Star product associated with NC-4 tori}\label{sec:star-products}
In this section, we will show how to construct a $*$-product equipped with which $C^{\infty}(\mathbb{T}^{4})$, the algebra of smooth functions on ordinary 4-torus turns into a noncommutative one. To this end, let us first write down the bijective Weyl map $\varpi:C^{\infty}(\mathbb{T}^{4})\rightarrow\mathcal{A}^{4}_{\theta(\rho,\sigma,\tau)}$ as
\begin{equation}\label{algebra-hom}
\varpi(\sum_{\bn\in\mathbb{Z}^{4}}f(\bn)e^{2\pi i\bn.\bx})=\sum_{\bn\in{\mathbb{Z}}^{4}}f(\bn)e^{\pi i\sum\limits_{j<k}n_{j}\theta_{jk}n_{k}}U^{\bn},
\end{equation}
where $\bn\equiv(n_1,n_2,n_3,n_4)\in\mathbb{Z}^{4}$, $f(\bn)\in\mathcal{S}(\mathbb{Z}^{4})$ and $U^{\bn}:=U_{1}^{n_1}U_{2}^{n_2}U_{3}^{n_3}U_{4}^{n_4}$ with $U_i$'s being the 4-unitary generators of $\mathcal{A}^{4}_{\theta(\rho,\sigma,\tau)}$. Here one uses the Fourier-expanded form of $f\in C^{\infty}(\mathbb{T}^4)$ written out as $f(\bx)=\sum\limits_{\bn\in\mathbb{Z}^{4}}f(\bn)e^{2\pi i\bn.\bx}$.

One immediately finds that $\varpi(fg)\neq\varpi(f)\varpi(g)$ for any $f, g\in C^{\infty}(\mathbb{T}^4)$. The bijective map $\varpi$ can be made into a homomorphism if one turns the commutative C* algebra $C^{\infty}(\mathbb{T}^4)$ into a noncommutative one by equipping it with the following $*$-product:
\begin{equation}\label{*-prod}
f*g:=\varpi^{-1}(\varpi(f)\varpi(g)),
\end{equation}
with $f, g\in C^{\infty}(\mathbb{T}^4)$.

One then obtains
\begin{eqnarray}\label{star-prod-comp}
\lefteqn{
\varpi(f)\varpi(g)}\nonumber\\
&&=\sum_{\bn,\mathbf{m}\in\mathbb{Z}^{4}}f(\bn)g(\mathbf{m})e^{\pi i\sum\limits_{j<k}(n_{j}\theta_{jk}n_{k}+m_{j}\theta_{jk}m_{k})}U^{\bn}U^{\mathbf{m}}\nonumber\\
&&=\sum_{\bn,\mathbf{m}\in\mathbb{Z}^{4}}e^{-\pi i\sum\limits_{j<k}(n_{j}\theta_{jk}m_{k}+m_{j}\theta_{jk}n_{k})}f(\bn)g(\mathbf{m})e^{\pi i\sum\limits_{j<k}(n_{j}+m_{j})\theta_{jk}(n_{k}+m_{k})}U^{\bn}U^{\mathbf{m}}\nonumber\\
&&=\sum_{\bn,\mathbf{m}\in\mathbb{Z}^{4}}e^{-\pi i\sum\limits_{j<k}(n_{j}\theta_{jk}m_{k}+m_{j}\theta_{jk}n_{k})}f(\bn)g(\mathbf{m})e^{\pi i\sum\limits_{j<k}(n_{j}+m_{j})\theta_{jk}(n_{k}+m_{k})}e^{2\pi i\sum\limits_{j<k}m_{j}\theta_{jk}n_{k}}U^{\bn+\mathbf{m}}\nonumber\\
&&=\sum_{\bn,\mathbf{m}\in\mathbb{Z}^{4}}e^{\pi i\sum\limits_{j,k=1}^{4}m_{j}\theta_{jk}n_{k}}f(\bn)g(\mathbf{m})e^{\pi i\sum\limits_{j<k}(n_{j}+m_{j})\theta_{jk}(n_{k}+m_{k})}U^{\bn+\mathbf{m}}\nonumber\\
&&=\sum_{\br,\bn\in\mathbb{Z}^{4}}e^{\pi i\sum\limits_{j,k}(r_{j}-n_{j})\theta_{jk}n_{k}}f(\bn)g(\br-\bn)e^{\pi i\sum\limits_{j<k}r_{j}\theta_{jk}r_{k}}U^{\br}\nonumber\\
&&=\sum_{\br\in\mathbb{Z}^{4}}\bigg[\sum_{\bn\in\mathbb{Z}^{4}}e^{\pi i\sum\limits_{j,k}r_{j}\theta_{jk}n_{k}}f(\bn)g(\br-\bn)\bigg]e^{\pi i\sum\limits_{j<k}r_{j}\theta_{jk}r_{k}}U^{\br},
\end{eqnarray}
so that one obtains
\begin{equation}\label{star-prod-express-right}
\varpi^{-1}(\varpi(f)\varpi(g))=\sum_{\br\in\mathbb{Z}^{4}}\bigg[\sum_{\bn\in\mathbb{Z}^{4}}e^{\pi i\sum\limits_{j,k}r_{j}\theta_{jk}n_{k}}f(\bn)g(\br-\bn)\bigg]e^{2\pi i\br.\bx}.
\end{equation}
But by using the Fourier mode expansion of the element $f*g\in C^{\infty}(\mathbb{T}^{4})$, i.e., $(f*g)(\bx)=\sum\limits_{\br\in\mathbb{Z}^{4}}(f*g)(\br)e^{2\pi i\br.\bx}$, one then concludes that
\begin{equation}\label{star-prod-final}
(f*g)(\br)=\sum_{\bn\in\mathbb{Z}^{4}}f(\bn)g(\br-\bn)e^{\pi i{\br}^{T}\theta\bn},
\end{equation}
where $\theta$ is the $4\times 4$ skew symmetric matrix provided by Theorem \ref{Theo:NC-tori}.
\begin{remark}\label{remark:twisted-group-Cstar-view}
A few remarks on the $*$-product obtained above in (\ref{star-prod-final}) are in order. Denoting $\sigma(\br,\bs):=e^{-\pi i\br^{T}\theta\bs}:\mathbb{Z}^{4}\times\mathbb{Z}^{4}\rightarrow\mathbb{T}$, one immediately observes that $\sigma(\br,\bs)$ is indeed a 2-cocycle on the Abelian group $\mathbb{Z}^{4}$. Therefore, for $\theta$ to be the skew-symmetric matrix given by Theorem \ref{Theo:NC-tori}, $\mathcal{A}^{4}_{\theta(\rho,\sigma,\tau)}$ can be defined as the twisted group C* algebra $C^{*}(\mathbb{Z}^{4},\sigma)$. At the classical limit when $\theta$ approaches the $0$-matrix, the $*$-product (\ref{star-prod-final}) reduces to the commutative convolution product making $\mathcal{A}^{4}_{\theta(\rho,\sigma,\tau)}$ the ordinary group C* algebra $C^{*}(\mathbb{Z}^{4})$ associated with the Abelian group $\mathbb{Z}^{4}$.
\end{remark}

\section{Projective modules over $\mathcal{A}^{4}_{\theta(\rho,\sigma,\tau)}$ and connections of constant curvature on them}\label{sec:classfctn-proj-mod}
Now that we have constructed the noncommutative differentiable manifold $\mathcal{A}^{4}_{\theta(\rho,\sigma,\tau)}$ explicitly out of the unitary dual $\gd$ in section \ref{sec:NC-tori-NCQM}, we proceed to construct  projective modules over such family of noncommutative C*-algebras. We follow the construction developed by M. Riefel in \cite{rieffel-class-proj} for this purpose. It is a known fact that projective modules over an important class of such noncommutative spaces, known as {\em irrational noncommutative tori} (see the definition in section \ref{sec:NC-tori-NCQM}) $\mathcal{A}^{n}_{\theta}$ are determined by the corresponding abelian $K_0$ group (see, for example, corollary (7.2) and theorem (7.3) of \cite{rieffel-class-proj} at p.258). As has been stressed by Rieffel in (p.289 of \cite{rieffel-class-proj}) that the Chern character of the elementary projective modules (to be explained in detail in this section) is the exact classification tool of the underlying projective modules unlike the canonical normalized trace which fails to be even faithful on the $K_0$ group of $\mathcal{A}^{n}_{\theta}$. From the pioneer work of Pimsner and Voiculescu (see \cite{Pimsner-Voiculescu}), it follows that the $K_0$ group of a noncommutative n-torus $\mathcal{A}^{n}_{\theta}$ is the same as that of an ordinary n-torus. In other words, one has $K_{0}(\mathcal{A}^{n}_{\theta})\cong\mathbb{Z}^{2^{n-1}}$. From which it follows that for the case of NC 4-tori we have constructed in the previous section, $K_{0}(\mathcal{A}^{4}_{\theta(\rho,\sigma,\tau)})\cong\mathbb{Z}^{8}$. 

Also, the building blocks for finitely generated projective modules over noncommutative $n$ tori are the {\em elementary projective modules} that arise from certain embedding maps. All other projective modules ({\em standard $\mathcal{A}^{n}_{\theta}$ modules} in the sense of p. 305, \cite{rieffel-class-proj}) can be obtained by tensoring the elementary projective modules with the finite dimensional projective representations of $\mathbb{Z}^{n}$ and then taking direct sum over such tensor product spaces. In the following section, we first construct the elementary projective right $\mathcal{A}^{4}_{\phi}$-module for some skew bilinear form $\phi$ on $\mathbb{Z}^{4}$ obtained from certain embedding map and then tensor them with finite dimensional cocycle representations of $\mathbb{Z}^{4}$ followed by rendering a projective right $\mathcal{A}^{4}_{\theta(\rho,\sigma,\tau)}$-module structure to the resulting tensor product space .

\subsection{Costruction of elementary projective modules over $\mathcal{A}^{4}_{\theta(\rho,\sigma,\tau)}$ and tensor products with finite dimensional representations}\label{subsec:classfctn-proj-module}
From the familiar construction of projective modules over higher dimensional noncommutative tori introduced by Rieffel (see \cite{rieffel-class-proj}), one knows that elementary projective modules over  are of the type $\mathcal{S}(\mathbb{R}^{p}\times\mathbb{Z}^{q})$ with $p$ and $q$ being nonnegative integers satisfying $2p+q=n$. Here $\mathcal{S}$ denotes the space of functions with rapid decay. For $n=4$, there are 3 types of elementary projective modules in the sense of \cite{rieffel-class-proj}:
\begin{eqnarray}
&\bullet& p=0, q=4: \mathcal{S}(\mathbb{Z}^{4})\label{eq:free-module}\\
&\bullet& p=2, q=0: \mathcal{S}(\mathbb{R}^{2})\\
&\bullet& p=1, q=2: \mathcal{S}(\mathbb{R}\times\mathbb{Z}^{2})
\end{eqnarray}
The modules given by (\ref{eq:free-module}) are free and hence can be written as a direct sum of finitely many copies of the underlying noncommutative 4-tori $\mathcal{A}^{4}_{\theta(\rho,\sigma,\tau)}$. We will be interested in constructing the other two nontrivial types of elementary projective modules explicitly over some $\mathcal{A}^{4}_{\phi}$ where $\phi$ is a bilinear form different from $\theta(\rho,\sigma,\tau)$. Following the construction of the elementary projective modules, we will tensor them with $\mathcal{S}(F)$ with $F$ being a finite commutative group and give the resulting tensor product space a projective $\mathcal{A}^{4}_{\theta(\rho,\sigma,\tau)}$-module structure. Note that our notation differs from that of Rieffel in that he used $\mathcal{A}_{\sigma}$-module instead of $\mathcal{A}_{\theta}$-module where the cocycle $\sigma$ is obtained by exponentiating the bilinear form $\theta$, i.e. $\sigma(x,y)=e^{-\pi i\theta(x,y)}$, with $x,y\in\mathbb{Z}^{4}$. 

For the case $p=1$ and $q=2$, the embedding map in Rieffel's notation is the map $T:\mathbb{R}^{4}\rightarrow\mathbb{R}^{3}\times\mathbb{R}^{*3}$. Therefore, it is conveniently represented as the $6\times 4$ matrix $T=(z_{ij})$ with $i=1,2,\ldots, 6$ and $j=1,2,\ldots,4$. Here, the Lie algebra of the 4-torus $\mathbb{T}^{4}$ is $\mathbb{R}^{4}$ and the lattice in the dual vector space $\mathbb{R}^{*4}$ is identified with $\mathbb{Z}^{4}$. Therefore, for each $x\in\mathbb{Z}^{4}$, the finitely generated projective module over the noncommutative 4-tori $\mathcal{A}^{4}_{\theta(\rho,\sigma,\tau)}$ should be given by an explicit action of operators $v_{x}$ on $\mathcal{S}(\mathbb{R}\times\mathbb{Z}^{2})\otimes\mathcal{S}(F)$ for some finite group $F$ according to Rieffel's general construction (see p. 288, \cite{rieffel-class-proj}). Here, we take the finite group $F$ to be $\mathbb{Z}_{M_{1}}\times\mathbb{Z}_{M_{2}}$ for given positive integers $M_{1}$ and $M_{2}$. For the sake of convenience, we will denote $v_{(1,0,0,0)}$, $v_{(0,1,0,0)}$, $v_{(0,0,1,0)}$ and $v_{(0,0,0,1)}$ by $v_1$, $v_2$, $v_3$ and $v_4$, respectively.

Now if one writes down the embedding $\mathbb{Z}^{4}\rightarrow\mathbb{R}^{3}\times\mathbb{R}^{*3}$ using the matrix $T$ introduced above as $x\mapsto(T^{\prime}(x),-T^{\prime\prime}(x))$, then using the general construction proposed by Rieffel in (p. 288, \cite{rieffel-class-proj}), the right $\mathcal{A}^{4}_{\phi(z_{ij})}$-module $\mathcal{S}(\mathbb{R}\times\mathbb{Z}^{2})$ can be given by
\begin{equation}\label{infinite-part-proj-mod}
(fv_{x})(m)=e^{2\pi i\langle m-T^{\prime}(x),T^{\prime\prime}(x)\rangle}f(m-T^{\prime}(x)),
\end{equation}
where $x\in\mathbb{Z}^{4}$ and $f\in\mathcal{S}(\mathbb{R}\times\mathbb{Z}^{2})$. Here, $\phi(z_{ij})$ is a skew-symmetric $4\times 4$ matrix with entries being quadratic polynomials in the entries $z_{ij}$ of the embedding map $T$ to be introduced shortly. In particular,  for generators $v_{i}$'s, with $i=1,\ldots,4$, one obtains,
\begin{equation}\label{infinite-part-gen-exprssn-proj-mod}
(fv_{i})(y,m_{1},m_{2})=e^{2\pi i(yz_{4i}+m_{1}z_{5i}+m_{2}z_{6i}-z_{1i}z_{4i}-z_{2i}z_{5i}-z_{3i}z_{6i})}f(y-z_{1i},m_{1}-z_{2i},m_{2}-z_{3i}).
\end{equation}

One then immediately finds that the following Weyl commutation relations hold:
\begin{equation}\label{weyl-comm-infinite-part}
v_{j}v_{k}=e^{-2\pi i\phi_{jk}}v_{k}v_{j},
\end{equation}
with each $\phi_{jk}$, for $j,k=1,\ldots,4$, is given by
\begin{equation}\label{noncomm-paramtr-infinite-part}
\phi_{jk}=-\begin{vmatrix}z_{1j}&z_{1k}\\z_{4j}&z_{4k}\end{vmatrix}-\begin{vmatrix}z_{2j}&z_{2k}\\z_{5j}&z_{5k}\end{vmatrix}-\begin{vmatrix}z_{3j}&z_{3k}\\z_{6j}&z_{6k}\end{vmatrix}.
\end{equation}
Now, $\phi_{jk}$'s, for $j,k=1,\ldots,4$, are precisely the entries of the skew-symmetric $4\times 4$ matrix $\phi(z_{ij})$ introduced earlier. One can immediately check using (\ref{noncomm-paramtr-infinite-part}) that $\phi(z_{ij})$ is indeed skew-symmetric. Suitable completion of $\mathcal{S}(\mathbb{R}\times\mathbb{Z}^{2})$ then gives an elementary projective right $\mathcal{A}^{4}_{\phi(z_{ij})}$-module that we will denote by $V^{T}$.

Now turning ourselves to the finite dimensional projective representations of $\mathbb{Z}^{4}$ on $\mathcal{S}(\mathbb{Z}_{M_1}\times\mathbb{Z}_{M_2})\cong\mathbb{C}^{M_{1}M_{2}}$ for positive integers $M_{1}$ and $M_{2}$, one can suitably choose a basis of $\mathbb{R}^{4}$, the Lie algebra of $\mathbb{T}^{4}$ (p. 321, \cite{rieffel-class-proj}) so that the matrices $w_{i}, i=1,\ldots,4$ representing the generators of $\mathcal{A}^{4}_{\psi(N_{i},M_{i})}$ act on the column vectors of $\mathbb{C}^{M_{1}M_{2}}$ in the following way (Here again $\psi(N_{i},M_{i})$ will denote a block-diagonalized skew-symmetric $4\times 4$ matrix with rational entries in positive integers $M_1$, $M_2$, $N_{1}\in\mathbb{Z}_{M_1}$ and $N_{2}\in\mathbb{Z}_{M_{2}}$):
\begin{equation}\label{finite-part-gen-exprssn-proj-mod}
\begin{split}
&(w_{1}\xi)(r,s)=\xi(\overline{r-N_{1}},s)\\
&(w_{2}\xi)(r,s)=e^{-\frac{2\pi ir}{M_{1}}}\xi(r,s)\\
&(w_{3}\xi)(r,s)=\xi(r,\overline{s-N_{2}})\\
&(w_{4}\xi)(r,s)=e^{-\frac{2\pi i s}{M_{2}}}\xi(r,s),
\end{split}
\end{equation}
where $(r,s)\in\mathbb{Z}_{M_1}\times\mathbb{Z}_{M_2}$ and the positive integers $N_{1}\in\mathbb{Z}_{M_1}$ and $N_{2}\in\mathbb{Z}_{M_2}$ are chosen to be relatively prime to $M_1$ and $M_2$, respectively. Also, by $\overline{r-N_{1}}$ and $\overline{s-N_{2}}$ we mean integer modulo $M_{1}$ and integer modulo $M_{2}$, respectively. Note also that, here, $\xi(r,s)$ denotes a component of the column vector $\xi\in\mathbb{C}^{M_{1}M_{2}}$. Therefore, $w_j$'s for $j=1,\ldots,4$ satisfy the following Weyl-commutation relations:
\begin{equation}\label{weyl-commut-finite-part}
w_{j}w_{k}=e^{-2\pi i\psi_{jk}}w_{k}w_{j},
\end{equation}
where $\psi_{jk}$'s, for $j,k=1,\ldots,4$, are the rational entries of the following skew-symmetric block-diagonalized $4\times 4$ matrix $\psi(N_{i},M_{i})$
\begin{equation}\label{skew-matrx-rational-part}
\psi(N_{i},M_{i})=\begin{bmatrix}0&-\frac{N_1}{M_1}&0&0\\\frac{N_1}{M_1}&0&0&0\\0&0&0&-\frac{N_2}{M_2}\\0&0&\frac{N_2}{M_2}&0\end{bmatrix}.
\end{equation}

\begin{remark}\label{remark-modular-arithmetic-calcultn}
A few remarks, on how to obtain the Weyl commutation relations (\ref{weyl-commut-finite-part}) from (\ref{finite-part-gen-exprssn-proj-mod}) with the $4\times 4$ skew-symmetric matrix $\psi(N_{i},M_{i})$ as given in (\ref{skew-matrx-rational-part}), are in order. For given $r,N_{1}\in\mathbb{Z}_{M_{1}}$ and $s,N_{2}\in\mathbb{Z}_{M_{2}}$ satisfying $r> N_{1}$ and $s>N_{2}$, $\overline{r-N_{1}}=r-N_{1}$ and $\overline{s-N_{2}}=s-N_{2}$ hold and (\ref{weyl-commut-finite-part}) follows from (\ref{finite-part-gen-exprssn-proj-mod}) using routine manipulation. Now, for given $N_{1}\in\mathbb{Z}_{M_{1}}$ and $N_{2}\in\mathbb{Z}_{M_{2}}$, if $r\in \mathbb{Z}_{M_1}$ and $s\in\mathbb{Z}_{M_2}$ are so chosen that $r<N_{1}$ and $s<N_{2}$ hold, then one immediately finds using $\frac{r-N_{1}}{M_{1}}=-1+\frac{\overline{r-N_{1}}}{M_{1}}$ that
\begin{equation*}
e^{2\pi i\left(\frac{\overline{r-N_{1}}}{M_{1}}\right)}=e^{2\pi i\left(\frac{r-N_{1}}{M_{1}}\right)}.
\end{equation*}
Similar arguments that hold for $s<N_{2}$ then lead one to the desired Weyl commutation relations (\ref{weyl-commut-finite-part}) from (\ref{finite-part-gen-exprssn-proj-mod}).
\end{remark}

Writing the operators acting on $\mathcal{S}(\mathbb{R}\times\mathbb{Z}^{2})\otimes\mathbb{C}^{M_{1}M_{2}}$ as $u_{j}=v_{j}\otimes w_{j}$, for $j=1,\ldots,4$, one immediately finds that for any $f\otimes\xi\in\mathcal{S}(\mathbb{R}\times\mathbb{Z}^{2})\otimes\mathbb{C}^{M_{1}M_{2}}$, the following holds
\begin{eqnarray}\label{total-Weyl-commutaion-derv}
((f\otimes\xi)u_{j})u_{k}&=&((f\otimes\xi)(v_{j}\otimes w_{j}))(v_{k}\otimes w_{k})\nonumber\\
&=&(fv_{j}\otimes\xi w_{j})(v_{k}\otimes w_{k})\nonumber\\
&=&(fv_{j})v_{k}\otimes(\xi w_{j})w_{k}\nonumber\\
&=&e^{2\pi i \phi_{kj}}(fv_{k})v_{j}\otimes e^{2\pi i\psi_{kj}}(\xi w_{k})w_{j}\nonumber\\
&=&e^{2\pi i(\phi_{kj}+\psi_{kj})}((f\otimes\xi)(v_{k}\otimes w_{k}))(v_{j}\otimes w_{j})\nonumber\\
&=&e^{2\pi i(\phi_{kj}+\psi_{kj})}((f\otimes\xi)u_{k})u_{j}.
\end{eqnarray}
In other words, for $j,k=1,\ldots,4$, one obtains,
\begin{equation}\label{total-Weyl-commt}
u_{j}u_{k}=e^{2\pi i(\phi_{kj}+\psi_{kj})}u_{k}u_{j}
\end{equation}
Using the defining relations of noncommutative tori (see (\ref{def:NC-tori})), one immediately finds that
\begin{equation}\label{total-nc-parameter}
\theta_{jk}=\phi_{jk}+\psi_{jk},
\end{equation} 
holds for $j,k=1,\ldots,4$ where $\theta_{jk}$'s are the entries of the $4\times 4$ skew-symmetric matrix $\theta(\rho,\sigma,\tau)$ given by Theorem \ref{Theo:NC-tori}. In matrix notation, (\ref{total-nc-parameter}) reads off as 
\begin{equation}\label{matrx-notatn-first-case}
\theta(\rho,\sigma,\tau)=\phi(z_{ij})+\psi(N_{i},M_{i}),
\end{equation}
so that the finitely generated $\mathcal{A}^{4}_{\phi(z_{ij})+\psi(N_{i},M_{i})}$-module $\mathcal{S}(\mathbb{R}\times\mathbb{Z}^{2})\otimes\mathcal{S}(\mathbb{Z}_{M_{1}}\times\mathbb{Z}_{M_2})$ is indeed an $\mathcal{A}^{4}_{\theta(\rho,\sigma,\tau)}$-module.
With $\theta(\rho,\sigma,\tau)$ given as above, one can define the right action of the operators $u_i$'s, for $i=1,\ldots,4$, on a given $g\in\mathcal{S}(\mathbb{R}\times\mathbb{Z}^{2})\otimes\mathcal{S}(\mathbb{Z}_{M_1}\times\mathbb{Z}_{M_{2}})$ in the following way
\begin{equation}\label{Total-acton-proj-mod}
\begin{split}
(gu_{1})(y,m_{1},m_{2},r,s)&=e^{2\pi i(yz_{41}+m_{1}z_{51}+m_{2}z_{61}-z_{11}z_{41}-z_{21}z_{51}-z_{31}z_{61})}\\
                                          &\times g(y-z_{11},m_{1}-z_{21},m_{2}-z_{31},\overline{r-N_{1}},s),\\
(gu_{2})(y,m_{1},m_{2},r,s)&=e^{2\pi i\left(yz_{42}+m_{1}z_{52}+m_{2}z_{62}-z_{12}z_{42}-z_{22}z_{52}-z_{32}z_{62}-\frac{r}{M_{1}}\right)}\\
                                          &\times g(y-z_{12},m_{1}-z_{22},m_{2}-z_{32},r,s),\\   
(gu_{3})(y,m_{1},m_{2},r,s)&=e^{2\pi i(yz_{43}+m_{1}z_{53}+m_{2}z_{63}-z_{13}z_{43}-z_{23}z_{53}-z_{33}z_{63})}\\
                                          &\times g(y-z_{13},m_{1}-z_{23},m_{2}-z_{33},r,\overline{s-N_{2}}),\\                                                                                
(gu_{4})(y,m_{1},m_{2},r,s)&=e^{2\pi i\left(yz_{44}+m_{1}z_{54}+m_{2}z_{64}-z_{14}z_{44}-z_{24}z_{54}-z_{34}z_{64}-\frac{s}{M_{2}}\right)}\\
                                          &\times g(y-z_{14},m_{1}-z_{24},m_{2}-z_{34},r,s),\\   
\end{split}
\end{equation}
where $(N_1,M_1)$ and $(N_2,M_2)$ are each pairs of relatively prime positive integers with $N_{i}\in\mathbb{Z}_{M_{i}}$ for $i=1,2$. The entries of the embedding map $T=(z_{ij})$ have to satisfy a system of 6 quadratic equations given by (\ref{total-nc-parameter}). Given such embedding map $T$ and $4$ positive integers $M_{1}$, $M_{2}$, $N_{1}\in\mathbb{Z}_{M_1}$ and $N_{2}\in\mathbb{Z}_{M_2}$ with $N_1$ and $N_2$ relatively prime to $M_1$ and $M_2$, respectively, one obtains precisely a projective right $\mathcal{A}^{4}_{\theta(\rho,\sigma,\tau)}$-module.

For the case, $p=2, q=0$, the finite dimensional projective representation of $\mathbb{Z}^{4}$ that one works with stays the same as before, i.e. $\mathcal{S}(\mathbb{Z}_{M_1}\times\mathbb{Z}_{M_2})$ with $M_1$ and $M_2$ being positive integers and hence the block-diagonalized skew-symmetric $4\times 4$ matrix $\psi(N_{i},M_{i})$ with rational entries is also given by (\ref{skew-matrx-rational-part}). However, the embedding map in this case changes to $\mathcal{T}:\mathbb{R}^{4}\rightarrow\mathbb{R}^{2}\times\mathbb{R}^{*2}$. In other words, it is represented by a $4\times 4$ matrix $\mathcal{T}=(z_{ij})$, with $i,j=1,\ldots,4$. Now following similar lines of arguments as in the previous case, one can write down the right action of the operators $v_i$'s with $i=1,\ldots,4$, representing the 4 generators of $\mathcal{A}^{4}_{\phi^{\prime}(z_{ij})}$ (Here, again, $\phi^{\prime}(z_{ij})$ is a skew-symmetric $4\times 4$ matrix with entries being quadratic polynomials in the entries $z_{ij}$ of $\mathcal{T}$ to be introduced shortly), on $\mathcal{S}(\mathbb{R}^{2})$ as
\begin{equation}\label{proj-exprssn-infinite-part-second-case}
(fv_{i})(y_{1},y_{2})=e^{2\pi i(y_{1}z_{3i}+y_{2}z_{4i}-z_{1i}z_{3i}-z_{2i}z_{4i})}f(y_{1}-z_{1i},y_{2}-z_{2i}),
\end{equation}
so that the Weyl commutation relations among these operators, for $j,k=1,\ldots,4$, are given by 
\begin{equation}\label{weyl-commutation-infinite-part-second-case}
v_{j}v_{k}=e^{-2\pi i\phi^{\prime}_{jk}}v_{k}v_{j},
\end{equation}
with each $\phi^{\prime}_{jk}$ now reads as follows
\begin{equation}\label{nc-parameter-second-case}
\phi^{\prime}_{jk}=-\begin{vmatrix}z_{1j}&z_{1k}\\z_{3j}&z_{3k}\end{vmatrix}-\begin{vmatrix}z_{2j}&z_{2k}\\z_{4j}&z_{4k}\end{vmatrix}.
\end{equation}
Here, $\phi^{\prime}_{jk}$'s are precisely the polynomial entries of the skew-symmetric $4\times 4$ matrix $\phi^{\prime}(z_{ij})$. It can be checked immediately from (\ref{nc-parameter-second-case}) that $\phi^{\prime}(z_{ij})$ is indeed a skew-symmetric matrix. By a suitable completion of $\mathcal{S}(\mathbb{R}^{2})$, one can thus obtain an elementary projective right $\mathcal{A}^{4}_{\phi^{\prime}(z_{ij})}$-module that we will denote by $V^{\mathcal{T}}$.

With the facts stated above, now, one can define the right action of the operators $u_i$'s on a given $g\in\mathcal{S}(\mathbb{R}^{2})\otimes\mathcal{S}(\mathbb{Z}_{M_1}\times\mathbb{Z}_{M_2})$, for $i=1,\ldots,4$ in the following way
\begin{equation}\label{proj-mod-gnrl-exprrsn-second-case}
\begin{split}
(gu_{1})(y_{1},y_{2},r,s)&=e^{2\pi i(y_{1}z_{31}+y_{2}z_{41}-z_{11}z_{31}-z_{21}z_{41})}g(y_{1}-z_{11},y_{2}-z_{21},\overline{r-N_{1}},s),\\
(gu_{2})(y_{1},y_{2},r,s)&=e^{2\pi i\left(y_{1}z_{32}+y_{2}z_{42}-z_{12}z_{32}-z_{22}z_{42}-\frac{r}{M_1}\right)}g(y_{1}-z_{12},y_{2}-z_{22},r,s),\\
(gu_{3})(y_{1},y_{2},r,s)&=e^{2\pi i(y_{1}z_{33}+y_{2}z_{43}-z_{13}z_{33}-z_{23}z_{43})}g(y_{1}-z_{13},y_{2}-z_{23},r,\overline{s-N_{2}}),\\
(gu_{4})(y_{1},y_{2},r,s)&=e^{2\pi i\left(y_{1}z_{34}+y_{2}z_{44}-z_{14}z_{34}-z_{24}z_{44}-\frac{s}{M_2}\right)}g(y_{1}-z_{14},y_{2}-z_{24},r,s),\\
\end{split}
\end{equation}
where, as in the previous case, $M_{1}$, $M_{2}$, $N_{1}\in\mathbb{Z}_{M_1}$ and $N_{2}\in\mathbb{Z}_{M_2}$ are all positive integers with $N_{1}$ and $N_{2}$ being relatively prime to $M_{1}$ and $M_{2}$, respectively. The operators, thus defined, obey the set of Weyl commutation relations given by 
\begin{equation}\label{Weyl-commutation-second-case}
u_{j}u_{k}=e^{-2\pi i\theta_{jk}}u_{k}u_{j}, 
\end{equation}
that satisfy
\begin{equation}\label{nc-parameters-relation-second-case}
\theta_{jk}=\phi^{\prime}_{jk}+\psi_{jk}, 
\end{equation}
where, for $j,k=1,\ldots,4$, $\psi_{jk}$'s are as in (\ref{skew-matrx-rational-part}) and each of $\phi^{\prime}_{jk}$'s, given by (\ref{nc-parameter-second-case}), is now expressed in terms of the entries of the matrix of the embedding map $\mathcal{T}$. Also, $\theta_{jk}$'s in (\ref{nc-parameters-relation-second-case}) can be read off from Theorem (\ref{Theo:NC-tori}) as entries of the skew-symmetric $4\times 4$ matrix $\theta(\rho,\sigma,\tau)$. In matrix notation, (\ref{nc-parameters-relation-second-case}) reads as
\begin{equation}\label{nc-parameter-matrix-notation-second-case}
\theta(\rho,\sigma,\tau)=\phi^{\prime}(z_{ij})+\psi(N_{i},M_{i}).
\end{equation}

 The entries of the $4\times 4$ matrix $\mathcal{T}$, like in the previous case, are again related by the $6$ quadratic constraints given by (\ref{nc-parameters-relation-second-case}) in terms of $\rho$, $\sigma$ and $\tau$ that label the unitary dual of the Lie group $\g$. The embedding map $\mathcal{T}:\mathbb{R}^{4}\rightarrow\mathbb{R}^{2}\times\mathbb{R}^{*2}$, whose entries are constrained to satisfy (\ref{nc-parameters-relation-second-case}), together with the above-mentioned  $4$ positive integers $M_1$, $M_2$, $N_{1}\in\mathbb{Z}_{M_1}$ and $N_{2}\in\mathbb{Z}_{M_{2}}$ determine a projective right $\mathcal{A}^{4}_{\theta(\rho,\sigma,\tau)}$-module.

\subsection{Construction of constant curvature connections on projective modules over $\mathcal{A}^{4}_{\theta(\rho,\sigma,\tau)}$}\label{subsec:constnt-curvature-connctn}
In the previous section \ref{subsec:classfctn-proj-module}, we constructed nontrivial elementary projective right $\mathcal{A}^{4}_{\phi(z_{ij})}$-module $V^{T}$ and $\mathcal{A}^{4}_{\phi^{\prime}(z_{ij})}$-module $V^{\mathcal{T}}$ with embedding maps $T$ and $\mathcal{T}$, respectively, using Rieffel's method described in \cite{rieffel-class-proj}. Subsequently, we tensored the above mentioned elementary projective modules with finite dimensional cocycle representations of $\mathbb{Z}^{4}$ and gave the resulting tensor product space a projective right $\mathcal{A}^{4}_{\theta(\rho,\sigma,\tau)}$-module structure.

The free module corresponding to $p=0$, there, is automatically projective and one can suitably define zero curvature connections on them (see, for example, p. 292, \cite{rieffel-class-proj}). In this section, we shall construct connections of constant curvatures on the projective right $\mathcal{A}^{4}_{\theta(\rho,\sigma,\tau)}$-modules constructed in section \ref{subsec:classfctn-proj-module}. 

Let $X$ be an element of the Lie algebra $\mathbb{R}^{4}$ of $\mathbb{T}^{4}$. As has been pointed out at the outset of section {\ref{subsec:classfctn-proj-module}} that $\mathbb{Z}^{4}$ is identified with the lattice in the vector space $\mathbb{R}^{*4}$. The dual pairing will be denoted by $\langle,\rangle$, henceforth. Now $X$ acts on the twisted group $C^{*}$ algebra $C^{*}(\mathbb{Z}^{4},\sigma)\equiv\mathcal{A}^{4}_{\theta(\rho,\sigma,\tau)}$ (see remark \ref{remark:twisted-group-Cstar-view} of section \ref{sec:star-products}) by means of the derivation $\delta_{X}$ in the following way:
\begin{equation}\label{derivtn-C-star-algbr}
\delta_{X}(u_{x})=2\pi i\langle X,x\rangle u_{x},
\end{equation}
for given $x\in\mathbb{Z}^{4}\subset\mathbb{R}^{*4}$.

Given $X\in\mathbb{R}^{4}$, the connection $\nabla_{X}$ on a projective right-module over a noncommutative 4-torus is expressed as an operator with the following action:
\begin{equation}\label{connctn-def}
\nabla_{X}(fu_{x})=(\nabla_{X}f)u_{x}+f(\delta_{X}(u_{x})),
\end{equation}
for some element $f$ lying in the underlying projective right-module over the given noncommutative 4-torus.

We now turn ourselves to the specific case $V^{T}$ of  elementary projective right $\mathcal{A}^{4}_{\phi(z_{ij})}$-module for the given embedding map $T$. Recall from section \ref{subsec:classfctn-proj-module} that for the case $p=1, q=2$, the embedding map was a map $T:\mathbb{R}^{4}\rightarrow\mathbb{R}^{3}\times\mathbb{R}^{*3}$ or more precisely $T:\mathbb{R}^{4}\rightarrow\mathbb{R}\times\mathbb{R}^{2}\times\mathbb{R}^{*}\times\mathbb{R}^{*2}$. Denoting by $\Pi$ the straightforward projection $\Pi:\mathbb{R}\times\mathbb{R}^{2}\times\mathbb{R}^{*}\times\mathbb{R}^{*2}\rightarrow\mathbb{R}\times\mathbb{R}^{2}\times\mathbb{R}^{*}$, one now finds that the composition $\widetilde{T}:=\Pi\mathrel{\circ}  T:\mathbb{R}^{4}\rightarrow\mathbb{R}\times\mathbb{R}^{2}\times\mathbb{R}^{*}$ is given by a $4\times 4$ matrix $\widetilde{T}=(z_{ij})$ just by deleting the last $2$ rows of the $6\times 4$ matrix $T$.

Now construct the following operators on $\mathcal{S}(\mathbb{R}\times\mathbb{Z}^{2})$ using the techniques adopted in (p. 290, \cite{rieffel-class-proj}) for given $r\in\mathbb{R}^{*}$, $(s,t)\in\mathbb{R}^{*2}$ and $w\in\mathbb{R}$ as
\begin{equation}\label{connection-oprtrs-first-case}
\begin{split}
&(Q^{1}_{r}f)(y,m_{1},m_{2})=(2\pi iyr)f(y,m_{1},m_{2})\\
&(Q^{2}_{(s,t)}f)(y,m_{1},m_{2})=2\pi i(m_{1}s+m_{2}t)f(y,m_{1},m_{2})\\
&(Q^{3}_{w}f)(y,m_{1},m_{2})=w\frac{\partial f}{\partial y}(y,m_{1},m_{2}).\\
\end{split}
\end{equation}
One now writes down the operator $Q_{(r,s,t,w)}$ as a sum of the above $3$ operators with $(r,s,t,w)\in\mathbb{R}^{*}\times\mathbb{R}^{*2}\times\mathbb{R}$. The action of $Q_{(r,s,t,w)}$ on $\mathcal{S}(\mathbb{R}\times\mathbb{Z}^{2})$ can then be read off as
\begin{equation}\label{Q-operator-first-case}
(Q_{(r,s,t,w)}f)(y,m_{1},m_{2})=2\pi i(yr+m_{s}+m_{2}t)f(y,m_{1},m_{2})+w\frac{\partial f}{\partial y}(y,m_{1},m_{2}).
\end{equation}
Now, for given $X\in\mathbb{R}^{4}$, the connection $\nabla_{X}$ on the elementary projective right $\mathcal{A}^{4}_{\phi(z_{ij})}$-module $V^{T}$ can be defined as
\begin{equation}\label{defin-conn-first-case}
\nabla_{X}=Q_{(\widetilde{T}^{-1})^{t}(X)}.
\end{equation}

If now one writes down $\widetilde{T}^{-1}=\frac{1}{\det\widetilde{T}}(C_{ji})$ or equivalently $(\widetilde{T}^{-1})^{t}=\frac{1}{\det\widetilde{T}}(C_{ij})$ using the cofactors $C_{ij}$'s of the $4\times 4$ matrix $\widetilde{T}$ for $i,j=1,\dots,4$, then $\nabla_{[\begin{smallmatrix}1&0&0&0\end{smallmatrix}]}$, $\nabla_{[\begin{smallmatrix}0&1&0&0\end{smallmatrix}]}$, $\nabla_{[\begin{smallmatrix}0&0&1&0\end{smallmatrix}]}$ and $\nabla_{[\begin{smallmatrix}0&0&0&1\end{smallmatrix}]}$, abbreviated as $\nabla_{1}$, $\nabla_{2}$, $\nabla_{3}$ and $\nabla_{4}$, respectively, can be explicitly expressed in terms of their actions on $\mathcal{S}(\mathbb{R}\times\mathbb{Z}^{2})$ in the following way:
\begin{equation}\label{connection-exprssn-first-case}
(\nabla_{j}f)(y,m_{1},m_{2})=\frac{2\pi i}{\det\widetilde{T}}(C_{j1}y+C_{j2}m_{1}+C_{j3}m_{2})f(y,m_{1},m_{2})+\frac{C_{j4}}{\det\widetilde{T}}\frac{\partial f}{\partial y}(y,m_{1},m_{2}),
\end{equation}
for $j=1,\ldots,4$. It is noteworthy that $C_{jk}$'s for $k=1,\ldots,4$ appearing in (\ref{connection-exprssn-first-case}) are cubic polynomials in the entries $z_{jk}$'s of $\widetilde{T}$ or the the entries of the first $4$ rows of the $6\times 4$ matrix $T$. But the entries of the embedding map $T$ are to satisfy the system of 6 quadratic equations given by (\ref{matrx-notatn-first-case}). In other words, the solution of the system (\ref{total-nc-parameter}) determines the cofactors $C_{jk}$'s for $k=1,\ldots,4$ involved in (\ref{connection-exprssn-first-case}).

Performing routine manipulation, one then obtains
\begin{equation}\label{curvature-exprssn-first-case}
[\nabla_{j},\nabla_{k}]=\frac{2\pi i}{(\det\widetilde{T})^{2}}(C_{j4}C_{k1}-C_{k4}C_{j1})\mathbb{I},
\end{equation}
where $\mathbb{I}$ is the identity operator on $\mathcal{S}(\mathbb{R}\times\mathbb{Z}^{2})$. One, therefore, verifies using (\ref{curvature-exprssn-first-case}) that the curvatures of the connections $\nabla_{j}$'s for $j=1,\ldots,4$ on $\mathcal{S}(\mathbb{R}\times\mathbb{Z}^{2})$, defined by (\ref{connection-exprssn-first-case}), are indeed constant, i.e. scalar multiple of the identity operator on the underlying elementary projective module. 

Let us remind ourselves that the connections are yet to be defined on $\mathcal{S}(\mathbb{R}\times\mathbb{Z}^{2})\otimes\mathcal{S}(\mathbb{Z}_{M_{1}}\times\mathbb{Z}_{M_{2}})$, i.e. projective right $\mathcal{A}^{4}_{\theta(\rho,\sigma,\tau)}$-module. In order to do so, let us recall from section \ref{subsec:classfctn-proj-module} that one defines the underlying tensor product for any $f\otimes\xi\in\mathcal{S}(\mathbb{R}\times\mathbb{Z}^{2})\otimes\mathbb{C}^{M_{1}M_{2}}$ as follows
\begin{equation}\label{tensor-prod-def}
(f\otimes\xi)(y,m_{1},m_{2},r,s)=f(y,m_{1},m_{2})\xi(r,s),
\end{equation}
where $\xi(r,s)$, as before, denotes a component of the column vector in $\mathbb{C}^{M_{1}M_{2}}$ and hence is represented by a complex number.

Now for a given $X\in\mathbb{R}^{4}$, define the connections $\widetilde{\nabla}_{X}$ on the projective right $\mathcal{A}^{4}_{\theta(\rho,\sigma,\tau)}$-module $\mathcal{S}(\mathbb{R}\times\mathbb{Z}^{2})\otimes\mathbb{C}^{M_{1}M_{2}}$ as follows
\begin{equation}\label{connection-def-full-mod}
\widetilde{\nabla}_{X}(f\otimes\xi)=\nabla_{X}f\otimes\xi.
\end{equation}

Then in terms of the $4$ connections $\nabla_{j}$'s as given by (\ref{connection-exprssn-first-case}), the connections $\widetilde{\nabla}_{j}$'s on $\mathcal{S}(\mathbb{R}\times\mathbb{Z}^{2})\otimes\mathbb{C}^{M_{1}M_{2}}$ can be written out for $j=1,\ldots,4$ as
\begin{eqnarray}\label{connection-action-full-first-case}
(\widetilde{\nabla}_{j}(f\otimes\xi))(y,m_{1},m_{2},r,s)&=&\frac{2\pi i}{\det\widetilde{T}}(C_{j1}y+C_{j2}m_{1}+C_{j3}m_{2})f(y,m_{1},m_{2})\xi(r,s)\nn\\
&+&\frac{C_{j4}}{\det\widetilde{T}}\frac{\partial f}{\partial y}(y,m_{1},m_{2})\xi(r,s).
\end{eqnarray}

Using (\ref{connection-def-full-mod}) together with (\ref{curvature-exprssn-first-case}) and (\ref{tensor-prod-def}), the curvature of the relevant connections can then be read off as
\begin{eqnarray}\label{curvature-full-first-case}
([\widetilde{\nabla}_{j},\widetilde{\nabla}_{k}](f\otimes\xi))(y,m_{1},m_{2},r,s)&=&\frac{2\pi i}{(\det\widetilde{T})^{2}}(C_{j4}C_{k1}-C_{k4}C_{j1})f(y,m_{1},m_{2})\xi(r,s)\nn\\
&=&\frac{2\pi i}{(\det\widetilde{T})^{2}}(C_{j4}C_{k1}-C_{k4}C_{j1})(f\otimes\xi)(y,m_{1},m_{2},r,s),\nn\\
\end{eqnarray}
for any $f\otimes\xi\in\mathcal{S}(\mathbb{R}\times\mathbb{Z}^{2})\otimes\mathbb{C}^{M_{1}M_{2}}$. In other words, one obtains the same constant curvature as in (\ref{curvature-exprssn-first-case}):
\begin{equation}\label{curvature-first-case-final}
[\widetilde{\nabla}_{j},\widetilde{\nabla}_{k}]=\frac{2\pi i}{(\det\widetilde{T})^{2}}(C_{j4}C_{k1}-C_{k4}C_{j1})\widetilde{\mathbb{I}},
\end{equation}
for $j,k=1,\ldots,4$ with $\widetilde{\mathbb{I}}$, now, denoting the identity operator on $\mathcal{S}(\mathbb{R}\times\mathbb{Z}^{2})\otimes\mathbb{C}^{M_{1}M_{2}}$.

Now consider the case $p=2, q=0$ where the embedding map $\mathcal{T}:\mathbb{R}^{4}\rightarrow\mathbb{R}^{2}\times\mathbb{R}^{*2}$ together with $4$ positive integers $M_{1}$, $M_{2}$, $N_{1}\in\mathbb{Z}_{M_{1}}$ and $N_{2}\in\mathbb{Z}_{M_{2}}$, with $N_{i}$ being relatively prime to $M_{i}$ for $i=1,2$, determine a projective right $\mathcal{A}^{4}_{\theta(\rho,\sigma,\tau)}$-module.

Repeating the construction preceding (\ref{defin-conn-first-case}) and using $\mathcal{T}^{-1}=\frac{1}{\det\mathcal{T}}(\mathcal{C}_{ji})$ or equivalently $(\mathcal{T}^{-1})^{t}=\frac{1}{\det\mathcal{T}}(\mathcal{C}_{ij})$, with $\mathcal{C}_{ij}$'s, for $i,j=1,\ldots,4$, being the cofactors of the $4\times 4$ matrix $\mathcal{T}$, one can neatly write down the connections $\nabla_{j}$'s on $\mathcal{S}(\mathbb{R}^{2})$ as
\begin{equation}\label{connection-second-case}
(\nabla_{j}f)(y_{1},y_{2})=\frac{2\pi i}{\det\mathcal{T}}(\mathcal{C}_{1j}y_{1}+\mathcal{C}_{2j}y_{2})f(y_{1},y_{2})+\frac{\mathcal{C}_{3j}}{\det\mathcal{T}}\frac{\partial f}{\partial y_{1}}(y_{1},y_{2})+\frac{\mathcal{C}_{4j}}{\det\mathcal{T}}\frac{\partial f}{\partial y_{2}}(y_{1},y_{2}),
\end{equation}
for $j=1,\ldots,4$.

The curvature associated with the connections (\ref{connection-second-case}) defined on $\mathcal{S}(\mathbb{R}^{2})$ can be read off as
\begin{equation}\label{curvature-second-case}
[\nabla_{j},\nabla_{k}]=\frac{2\pi i}{(\det\mathcal{T})^{2}}(\mathcal{C}_{3j}\mathcal{C}_{1k}+\mathcal{C}_{4j}\mathcal{C}_{2k}-\mathcal{C}_{3k}\mathcal{C}_{1j}-\mathcal{C}_{4k}\mathcal{C}_{2j})\mathcal{I},
\end{equation}
for $j,k=1,\ldots,4$ and $\mathcal{I}$ being the identity operator on $\mathcal{S}(\mathbb{R}^{2})$.

Along the same line of arguments as presented in the previous case for $p=1,q=2$, if one now defines a connection on $\mathcal{S}(\mathbb{R}^{2})\otimes\mathbb{C}^{M_{1}M_{2}}$ for a given $X\in\mathbb{R}^{4}$ and $f\otimes\xi\in\mathcal{S}(\mathbb{R}^{2})\otimes\mathbb{C}^{M_{1}M_{2}}$ as
\begin{equation}\label{full-crvature-def-second-case}
\widetilde{\nabla}_{X}(f\otimes\xi)=\nabla_{X}f\otimes\xi,
\end{equation}
then one can immediately compute the curvature pertaining to these connections to be
\begin{equation}\label{final-exprssn-curvature-second-case}
[\widetilde{\nabla}_{j},\widetilde{\nabla}_{k}]=\frac{2\pi i}{(\det\mathcal{T})^{2}}(\mathcal{C}_{3j}\mathcal{C}_{1k}+\mathcal{C}_{4j}\mathcal{C}_{2k}-\mathcal{C}_{3k}\mathcal{C}_{1j}-\mathcal{C}_{4k}\mathcal{C}_{2j})\widetilde{\mathcal{I}},
\end{equation}
with $\widetilde{\mathcal{I}}$ denoting the identity operator on $\mathcal{S}(\mathbb{R}^{2})\otimes\mathbb{C}^{M_{1}M_{2}}$.

Therefore, we find that the connections, defined by (\ref{connection-second-case}) on $\mathcal{S}(\mathbb{R}^{2})\otimes\mathbb{C}^{M_{1}M_{2}}$, are indeed of constant curvature, i.e. scalar multiple of the identity operator on the respective module. Here, the solution of the system of quadratic equations (\ref{nc-parameter-matrix-notation-second-case}) determines the cofactors $\mathcal{C}_{kj}$'s for $k=1,\ldots,4$ involved in (\ref{connection-second-case}).

\section{Conclusion and future perspectives}\label{sec:conclsn-perspctve}
In this paper, we have demonstrated that there exists a 2-parameter family of vector potentials associated with a 2-dimensional quantum system in a constant external magnetic field. How these choices of gauge potentials are related with the unitarily equivalent irreducible representations of $\g$ is subsequently discussed along with specific examples of the widely used gauges in the literature. Later, we constructed a noncommutative $4$-tori out of the unitary dual of $\g$. Star-product between smooth functions on the $4$-torus $\mathbb{T}^{4}$ is subsequently constructed. Noncommutative tori are the most widely studied examples of noncommutative differentiable manifolds that have extensive applications in Mathematical Physics, such as discrete Schr\"odinger operators with almost periodic potentials, the almost Mathieu operators and quantum diffusions (consult \cite{rieffel} for more detail). In the present case of noncommutative $4$-tori obtained from the unitary dual of $\g$, we constructed finitely generated projective modules over them and subsequently defined connections of constant curvature on such projective modules. 

Wigner functions associated with various coadjoint orbits of $\g$ and the star-products between them are constructed in \cite{wigfuncpaper}. It will be interesting to see how these Wigner functions turn out to be for the other gauge equivalent representations (see (\ref{gauge-equiv-irrep})) of $\g$. We wish to define equivalence relations between the star-products emerging from equivalent unitary irreducible representations of $\g$ and study their possible relationship with the existing literature (see, for example, \cite{gutetalt}). Another interesting idea would be to study the spectrum of the Hamiltonian of {\em noncommutative harmonic oscillator} due to the representation (\ref{gauge-equiv-reps-algbr}) of $\G$ and investigate the roles of the $2$ continuous parameters $l$ and $m$ here. We also wish to study other noncommutative geometrical aspects associated to NCQM, for example,  study spin geometry for$\mathcal{A}^{4}_{\theta(\rho,\sigma,\tau)}$ and compute Dirac operator on the associated spinor module. We will also be keen to know when two algebras from the family $\mathcal{A}^{4}_{\theta(\rho,\sigma,\tau)}$ constructed in section \ref{sec:NC-tori-NCQM} become Morita equivalent to each other in terms of the parameters $\rho$, $\sigma$ and $\tau$ that label the unitary dual of $\g$.

\section{Acknowledgements}
The author is thankful to the referee for many useful comments. He also expresses his heartfelt gratitude to his mentor late professor Syed Twareque Ali to help him choose an appropriate title for the manuscript. Many thanks also go out to professor Arshad Momen (Dhaka University) and professor Hishamuddin Zainuddin (INSPEM, UPM) for stimulating discussions. He also acknowledges a grant from National Natural Science Foundation of China (NSFC) under Grant No. 11550110186.

\end{document}